\documentclass[10pt,letterpaper]{article}
\usepackage[top=0.85in,left=2.75in,footskip=0.75in]{geometry}

\usepackage{amsmath,amssymb}

\usepackage{changepage}

\usepackage[utf8x]{inputenc}

\usepackage{textcomp,marvosym}

\usepackage{cite}

\usepackage{nameref,hyperref}

\usepackage[right]{lineno}

\usepackage{microtype}
\DisableLigatures[f]{encoding = *, family = * }

\usepackage[table]{xcolor}

\usepackage{array}

\usepackage[nomarkers,figuresonly]{endfloat}

\newcolumntype{+}{!{\vrule width 2pt}}

\newlength\savedwidth

\raggedright
\usepackage[aboveskip=1pt,labelfont=bf,labelsep=period,justification=raggedright,singlelinecheck=off]{caption}

\makeatletter
\renewcommand{\@biblabel}[1]{\quad#1.}
\makeatother

\usepackage{lastpage,fancyhdr,graphicx}
\usepackage{epstopdf}
\fancyheadoffset[L]{2.25in}
\fancyfootoffset[L]{2.25in}
\begin{document}
\vspace*{0.2in}

\begin{flushleft}
{\Large
\textbf\newline{\textbf{Simple estimators for network sampling}} % Please use "sentence case" for title and headings (capitalize only the first word in a title (or heading), the first word in a subtitle (or subheading), and any proper nouns).
}
\newline
\\
Steve Thompson\textsuperscript{1*}
\\
\bigskip
\textbf{1} Department of Statistics and Actuarial Science, Simon
Fraser University, Burnaby, British Columbia, Canada
\\

\bigskip

% Insert additional author notes using the symbols described below. Insert symbol callouts after author names as necessary.
% 
% Remove or comment out the author notes below if they aren't used.
%
% Primary Equal Contribution Note
%\Yinyang These authors contributed equally to this work.

% Additional Equal Contribution Note
% Also use this double-dagger symbol for special authorship notes, such as senior authorship.
%\ddag These authors also contributed equally to this work.

%\textcurrency Current Address: Dept/Program/Center, Institution Name, City, State, Country % change symbol to "\textcurrency a" if more than one current address note
% \textcurrency b Insert second current address 
% \textcurrency c Insert third current address


% Group/Consortium Author Note
%\textpilcrow Membership list can be found in the Acknowledgments section.

% Use the asterisk to denote corresponding authorship and provide email address in note below.
* thompson@sfu.ca

\end{flushleft}

\section*{Abstract}

A new estimation method is presented for network sampling designs,
including Respondent Driven Sampling (RDS) and Snowball (SB) sampling.  These types of link-tracing designs  are essential for studies of
hidden populations,  such as people at risk for HIV.  The simple
idea behind the new method is to run a fast-sampling
process on the sample network data to estimate the inclusion
probabilities of the actual survey, and incorporate those in unequal
probability estimators of population means and proportions.  

Improved versions of the usual RDS and SB designs are also proposed,
termed RDS+ and SB+, to obtain information on more of the
within-sample links.  In simulations using the network from the
Colorado Springs study on the heterosexual spread of HIV, the new
estimators produce in most cases lower bias and lower mean square than
current methods. For the 
variables having the largest mean square errors with current
estimators, the improvement with the new estimator is dramatic.  The
estimates are improved even more with the enhanced design versions.
For estimating the population mean degree, the efficiency gains using
he new method are 29 for RDS, 54 for RDS+, 26 for SB and 80 for SB+.
This means for example, with the ordinary RDS design, the mean square
error with the new estimator, same data, is 1/29 that of currently
used estimators. 

The new method is computationally intricate but is fast and scales up
well.  The new estimation method can be used to
re-analyze existing network survey data.  For new network sampling
studies, it is recommended to use  the improved
designs as well as the new estimators.

% Please keep the Author Summary between 150 and 200 words
% Use first person. PLOS ONE authors please skip this step. 
% Author Summary not valid for PLOS ONE submissions.   
\section*{Author summary}

This paper presents new estimators of characteristics of hidden and
hard-to-reach populations using data from network sampling designs.
Network sampling designs such as Respondent-Driven Sampling and
snowball sampling are important for studying hidden populations
including key populations with high exposure to the HIV epidemic
through drug injecting or sexual links.   These surveys are essential for
the understanding needed to bring relief to those affected and to reduce
or end the epidemic.

Evaluations of the new estimators are carried out using data from a
large-scale network study of a high-risk population.  The new method
produces more accurate estimates of population values in most cases
compared to currently used methods.  The increase in accuracy is
achieved by eliminating most of the bias.   The reason for the better
performance of the new estimators is that they do not rely on unrealistic
assumptions used by other estimators about how the sample is
selected in the real world.  Instead, the new estimator uses a fast
sampling process, similar to the real-world design, on the sample
network data to estimate the relative probability of each person being
included in the sample.  These inclusion probability estimates are
used in estimates of population characteristics.

%\linenumbers

\section*{Introduction}

Network sampling consists of designs which at least in part follow
network links from units already in the sample to add connected units
to the sample.  The fundamental problems in network sampling are to
find the best designs for a given situation and then, given the sample
data, to find effective inference methods for estimating
characteristics of the population.  Network sampling methods have long
been used to investigate hidden and hard-to-reach populations, and in
many cases provide the only feasible means to sample such a
population.  In this paper I introduce new simple estimators for
inference from network samples that work for a variety of network
sampling designs including the types most widely used.  

The HIV/AIDS pandemic has provided motivation for the development and
use of network sampling methods.  The most recent UNAIDS report
\cite{unaids2018report} estimates that 36.9 million people worldwide
are living with HIV, with 1.8 million new infections and 940,000
AIDS-related deaths annually.  Academic work and public funding of
research into network sampling methods for key populations at risk
have increased dramatically since the late 1990s.  A recent systematic review
\cite{white2015strengthening} found over 460 peer-reviewed papers on
Respondent Driven Sampling surveys, not including methodology studies,
in 69 countries worldwide.  Another recent paper  \cite{verdery2015network} reported
finding 642 academic articles on Respondent-Driven Sampling.  A search of the NIH funded grant database
in 2012 
\cite{mouw2012network} found that over \$100 million had been
provided for grants with ``Respondent-Driven Sampling'' in the title
or key words, while by 2015 the figure exceeded \$180 million
\cite{verdery2015network} in NIH funding.  Many of the actual RDS studies have been funded by the
U.S. President's Emergency Plan for AIDS Relief (PEPFAR) and the
Centers for Disease Control and Prevention. 

The potential benefit of these network sampling surveys in these key
populations is huge.  These link tracing methods provide the means to
get into these hard to reach populations, to understand the biological
and behavioral factors that enhance or inhibit the virus. Most of
these surveys include a bioassay component, with saliva or blood
samples taken, tested for HIV and other infections such as HSV2, and
sometimes sequenced for HIV strain.  In addition, the link-tracing
samples find there way into the parts of the population where the
spread of the virus is most explosive and hard to eradicate.  These
surveys bring with them interventions, such as HIV testing, referral
for antiretroviral prescriptions, counseling, and condoms, to the
people most in need of them.  The understanding from these surveys is
central to efforts to contain and bring down the epidemic globally.
Estimates from these surveys provide essential input for simulations
of the HIV epidemic and evaluations of intervention programs.

The simple idea behind the new estimates is to run a fast-sampling
process, similar to the actual sampling design, on the sample network
data, and use the fast-sample inclusion frequencies to estimate the
unknown survey-inclusion probabilities and thereby estimate the
population values.  In effect the new method explores computationally
the sample network topology to full full depth and weights each
observation using all the network paths leading to that node,
path-lengths affecting weights and redundant pathways lending
increased weight.  In contrast, the most widely used current methods
either do not use the sample network data at all, using self reported
degree instead, or using only the first-step depth of the sample
network data by comparing numbers of connections within and between
sets of nodes.

For computation there are two approaches to implementing the simple
idea.  One is to sample the network data repeatedly  with a
link-tracing design similar to the original design, from seed
selection to target sample size.  Because the
design is without replacement, a smaller sample size needs to be
used than was used in the original design.  In this approach a sequence independent samples are selected,
each one proceeding from seeds to target sample size.  The second
approach is to construct a sampling process that selects a sequence of
samples that are not independent but have selection properties 
similar to the original design.  These approaches are described in
detail in the Methods section.

The inference method described in this paper uses data from network
sampling designs such as Respondent Driven Sampling (RDS) and Snowball
(SB) designs and produces estimates and confidence intervals for
population means and proportions of the survey response variables.
No assumptions are made about the population network.  Instead,
information about the network is extracted from the sample network
data, that is, the data on links between sample nodes.  From the
sample network data the relative inclusion probabilities of the sample
nodes are estimated, and design-based estimates of population values
are made from those.  Improved versions of the RDS and SB designs,
termed RDS+ and SB+, are also proposed in this paper.  The improved designs
increase the number of sample links in the data, resulting in better
estimates for little additional effort in data collection.

The basic difficulty for inference from link-tracing sampling designs
is that the sampling procedure selects different units with different
probabilities, depending on the unit's place among the links of the
population network.  A widely used and much-studied estimator for use
with unequal probability designs is the generalized unequal
probability estimator (GUPE) \cite{brewer1963ratio}, which divides
each sample unit's value by the unit's inclusion probability and
normalizes by the sample sum of the reciprocals of the inclusion
probability.  With most of the network sampling designs that are
actually used for hidden populations, the inclusion probabilities are
unknown and depend on network links beyond the sample data.

In practice modern network sampling designs for hidden populations
work by handing out coupons to each person recruited into the sample,
with which they can recruit people to whom they are linked.  When any
person comes in with a coupon, they are paid a small honorarium, as
is the person who gave them the coupon.  After being interviewed and
in some cases undergoing medical tests, the new recruit is in turn
given coupons with which to recruit new sample members.  This process
continues until target sample size is met.  With RDS designs the
number of coupons is usually limited to a small number such as 2 or
three.  With snowball designs the number of coupons given equals the
number of partners reported, or else a high coupon limit is set, such
as 15 or 25.   The survey is started out with an initial sample, or
``seeds'' selected by survey personnel.   

For a one-wave snowball design \cite{frank1994estimating} offered a number
of design-based and model-based approaches to estimating the size of a
hidden population.  For RDS surveys, Salganik and Heckathorn (SH)
\cite{salganik2004sampling} introduced an estimator for mean degree
which uses the form of the GUPE but with degree $d_i$ of node $i$
replacing the inclusion probability $pi_i$.  This was based on the
assumption that the recruitment process by which the sample was
selected was similar to a with-replacement random walk, producing a
Markov chain in which the current state of the chain is the currently
selected node and on the assumption that the population network
consisted of only a single connected component.  Using a separate
assumption that the sequence of selections having the attribute and
not having the attribute was also a Markov chain they arrived at an
estimator of the proportion of the population having the attribute.
That estimator, in addition to using the self-reported degrees, uses
the proportion of recruitment links leading out from a group that go
to the other group.  In this way, their estimator uses information
from the sample network link data, in addition to the sample node data
such as attribute value and degree of a node.  Heckathorn
\cite{heckathorn20076} extends that method to estimate the several
proportions of a categorical variable and to estimate the group mean
of a continuous variable.

The Volz-Heckathorn estimator (VH) \cite{volz2008probability}, using
the random walk approximation for the sampling design and assuming a
single component in the population, uses the form of the GUPE with
degree in place of inclusion probability for estimates of means of all
types of node variables, whether attribute or more general numeric.
This estimator does not use any of the recruitment link or other
sample network data, only the node variable of interest and the node
reported degree.  The VH estimator is currently the most commonly used
estimation method used with RDS surveys.

The Successive Sampling estimator of Gile \cite{gile2011improved}
improves on the VH estimator for samples in which sample size $n$ is a
substantial fraction of population size $N$.  The improvement is based
on the fact that the real sampling is done without replacement.  The
effect of the sampling fraction $n/N$ is adjusted for using a
stochastic sampling algorithm to estimate the inclusion probabilities
with a successive sampling without-replacement design in which at the
$i$-th selection a node is selected with probability proportional to
degree from among all units not already selected in the first $i-1$
steps.  This procedure involves estimating iteratively the unit
degrees in the whole population of size $N$, and therefore it requires
knowledge or an estimate of population size $N$.  The estimated
inclusion probabilities $\hat \pi_i$ are then used in the GUPE form to
estimate population means and proportions, for any type of node
variable.  When sample fraction $n/N$ is small, this estimator
defaults to the VH estimator.  A simple network model called the
configuration graph model, which takes any arbitrary fixed degree
distribution and randomly connects the link ends, is used to motivate
the SS model.

Fellows \cite{fellows2018respondent} introduces the Homophily
Configuration Graph model, which for an arbitrary degree distribution
allows the linkage probabilities to depend on attribute memberships.
This model motivates the Homophily Configuration Graph Estimator
(HCG), which uses the attribute variable group membership in a manner
similar to SH and adjusts for large sampling fraction with the SS
method.  The method uses recruitment link data as in the SH and
requires an estimate $\hat N$ as in SS.  Fellows shows that the SH
method works well under the configuration graph model and that HCG
estimator works well under the homophily configuration graph model.
Simulations show the method works well under homophily configuration
graph model realizations as expect.  It also works well for the
attribute variables of the empirical Project 90 network, which is also
used in the present paper, though he uses for the simulation only the
largest connected component from that network.  The method is
described for estimating the mean of an attribute variable, which is
the population proportion of that variable, as uses the estimate from
the recruit links of transitions between the groups with the attribute
and without.

Confidence interval methods commonly used with RDS designs include the
Salganik bootstrap \cite{salganik2006variance} for SH and VH, the Gile
SS bootstrap \cite{gile2011improved} for SS.  A recent evaluation of
these methods, for means of binary variables and using simulations
based on a statistical network model fitted to RDS data is
\cite{spiller2017evaluating}.  A different bootstrap approach for RDS
  data called a Tree Bootstrap is described in
  \cite{baraff2016estimating}.

Estimation methods based on the VH estimator are based on node
degrees.  The SH estimator goes one step further into the sample
network by using the proportion of links within a group to the number
going out from that group.  In contrast the methods in \cite{thompson2006aws},
\cite{crawford2016graphical}, and \cite{crawford2018hidden} use the
full sample network.  Minimally, the sample network includes only
those links within the sample that were used in recruitment, plus the
counterpart edge in the other direction to symmetrize the link.  The
more full version of the sample network uses the set of all links that
connect sample nodes.  

\cite{thompson2006aws} requires the full sample subgraph in order to
apply the Rao-Blackwell method to improve an initial unbiased
estimator.  \cite{crawford2016graphical} gives a method for estimating
the sample subgraph from the recruitment graph, and
\cite{crawford2018hidden} uses that estimated subgraph, together with
an assumed model, to estimate the size of a hidden population.  The
method proposed here uses whatever sample network information is
available.  If that is only the recruitment graph, that alone is used
in estimation.  No attempt is made to estimate the sample subgraph.
Estimation efficiency is improved if more within-sample links are
known.  These are the enhanced versions of the estimators.  The best
estimators result from designs which reveal all the within-sample
links.

The methods such as VH, SH, SS, and HCG that use underlying Markov
chain assumptions about the sampling design address estimation only
with RDS type of designs using small numbers of coupons per
participant, such as 2 or 3, limiting the number of recruits each
recruiter can bring in.  More general snowball types of designs, which
issue as many coupons as the reported number of partners or sets a
high maximum coupon limit such as 15 or 25, have not been addressed or
evaluated for those methods.  Each of the methods above, SH, VH, SS,
and HCG uses in part the reciprocal of reported degree based on an
approximating random walk assumption, so that each is related to the
VH estimator.  For estimating mean degree, SH defaults to VH.  With
large population size in relation to sample size, SS defaults to VH.
And for estimating mean degree with large population relative to
sample size, HCG defaults to VH.

The new method of inference proposed here works for any of the
commonly used network sampling designs.  This includes the
snowball-like designs as well as the small-coupon-number
designs.  The new method estimates the population means or proportions of all
types of variables in the same way, whether they are continuous,
integer, or binary valued.   It does not assume a network model producing
networks with particular properties, nor does it assume the population
network contains only one connected component.  It does not assume the
sampling design is similar to a with-replacement random walk design,
nor that population values of variables of interest (node $y$-values)
have any particular pattern or that the selected sequence of those
values has Markov-chain properties.  

Instead, the new method uses the sample network data and examines
computationally its network topology.  That is, it examines the
patterns of connectedness in the links part of the sample $E_s$ and
with a fast sampling process calculates the inclusion frequencies of
those units in a network sampling design.  The inclusion probability
of a node $i$ is high if a lot of paths lead to it from other nodes.
Multiple or redundant paths to it increase it's probability of being
included.  The probabilities are influenced by the lengths of the
different paths and by the inclusion probabilities of the other nodes
on those paths.  Further, the exact inclusion probabilities for units
in the sample depend on sample paths that come in from with wider
network outside of the sample.  So a fast sampling process restricted
to the sample network, using link-tracing similar to the original
design, is used.  The long-term relative inclusion frequencies with the
fast-sampling process are used as estimates of the relative inclusion
probabilities.  And these estimated inclusion probabilities are used to
estimate the population characteristics.  

The method is computationally efficient and scales up extremely well
to larger sample sizes.   In the simulations the new method gives
better estimates of population values in most cases, compared to the
compared to other estimators, and in some
important cases the estimates are many times better with the new
method.  The estimates are in most cases better still with the
improved improved (``plus'') versions of the network sampling
designs. 

In the empirical simulations of this paper the new estimators perform
in most cases better than the VH estimator, in terms of bias and mean
square error.  In some important cases, such as for estimating
population mean degree, the new estimators perform very much better.
The substantial bias of the VH estimator arises largely from the the
discrepancy between node degree and the actual inclusion probability
of a node.  This discrepancy in turn arises from the difference
between an assumed with-replacement random walk in a strongly
connected graph and the actual recruitment process used in surveys of
hidden populations.  The actual survey design is without-replacement,
and with more than one coupon the selection process branches.  The
population network tends to have more than one component rather than
being strongly connected.  \cite {goel2010assessing} and
\cite{gile20107} call attention to the biases arising from these
unrealistic assumptions.  The methods of \cite{gile2011improved} and
\cite{fellows2018respondent} serve to reduce the part of the
without-replacement bias for the case of large sampling fractions.
Even with a relatively small sample fraction as in the simulations of
this paper, the without-replacement sampling and branching pattern
produce marked differences from the random walk assumptions.  While
the bias of the VH estimator is higher with the snowball designs than
with the small-coupon-number designs, it is large with both.

The new estimators do not rely on assumptions about Markov chain
properties of the recruitments or transitions between respondent
values, nor do they depend on the network having only one connected
component.  Instead, the new estimators explore the sample network
data empirically.  The exploration is done by selecting a sequence of
samples from the network data using a fast sampling process having
features similar to the real sample selection design.  These features
include branching, without-replacement selections, and a a small rate
or reseeding to be sure the sample does not get stuck in any single
connected component.

By using the fast process that is similar to the real design and
exploring with it the paths of all lengths reaching reaching to and
from each unit, the new estimators get more accurate estimates of the
inclusion probabilities, which greatly reduces the bias of the
estimates of population characteristics.  Existing network survey
data can be reanalyzed to obtain better estimates using the new
estimators.  For future surveys, the proposed data collection
enhancements which provide information on additional within-sample
links produce further estimation improvements.  Considering the costs
and benefits of these surveys worldwide, these improvements are well
work pursuing.

\section*{Network sampling designs}

A network sample has the form 
\begin{equation}
s = (U_s, E_s)
\end{equation} 
where $U_s$ denotes the units, or nodes, in the sample and $E_s$ denotes the
edges, or links, in the sample.  (``Nodes'' and ``Units'' are used
interchangeable, as are ``edges'' and ``links'' in this paper.)  Edges
may be directional or symmetric.

A network sampling design is a procedure for getting such a sample
from a population.  Usually edges are discovered through observing or
interviewing sample nodes, and edges may be traced to add more nodes
to the sample.  The traced edges become part of the edge part of the
sample.  Additional within-sample edges, beyond those used for
recruitment,  can also be added to added to the sample by various
means.  

Two types of network sampling designs in current use for
hard-to-access human populations include Respondent Driven Sampling
(RDS) and Snowball Sampling (SB).  In RDS, 1.  a sample of initial
units called ``seeds'' are selected by some means;  2.  each member of
the sample is given a small number of coupons, like two or three, with
which to recruit new people who are connected to the sample member by
the type of edge relationship of interest;  3.  Those recruits who
come in are themselves given coupons with which to recruit additional
people, and sampling continues until a target sample size is reached.

In Snowball Sampling designs, 1.  a sample of seeds is selected;  2.
each sample member is given as many coupons as their number of
partners, or up to some high limit such as 15 or 25;  3.  sampling
continues until target sample size is reached.

The sample of nodes in these network designs applied to human
populations includes the seeds and the people recruited in the
subsequent waves of sampling.  The data collected on these sample
members includes interview responses on demographics and behavior and
biological assay results such as results of blood or saliva tests for
infections.  

The sample of links includes all the relationships known between
sample members.  In ordinary RDS and Snowball sampling protocols, the
only links known sample members are those links used in recruitment.
So $E_s$ consists only of recruitment links.  In this paper we
consider also two enhancements of these designs, termed RDS Plus In
RDS plus, an RDS sample of nodes is selected and extra effort is made
to obtain information on as many of the other links between sample
members.  Similarly, with the SB Plus design, additional effort is
made to find the additional within-sample links.  The Plus version of
each design type obtains the same sample $U_s$ of nodes, but a larger
sample $E_s$ of links.

A number of procedures have been used in surveys for discovering the
additional within-sample links.  For instance, 1.  multiple
recruitments of a person by different recruiters can be allowed in he
protocol, revealing additional within-sample links without having to
do repeat interviews.  2.  Unique identifiers of partners can be
obtained and made anonymous but revealing more links than revealed
through recruitments.  An example of such a procedure is asking each
sample member for names or phone numbers of partners, as well as their
own name or phone numbers.  The names or phone numbers can be changed
to code values and the code values matched up to provide sample links.
An effective anonymizing system for such unique identifiers is
described by Fellows (2012) \cite{fellows2012exponential}.  3.
Intensive in-community ethnographic or epidemiological field work can
trace out and identify the individuals at both ends of partnership
links, as has been done in studies of hard to access populations such
as the Project 90 study providing the data for the evaluation
simulations in this paper.  The field ethnographic methods identify
names of partners, with the actual names later being removed from the
data, replaced by arbitrary ID numbers informative only within the
data set.

\section*{Fast sampling process}

Given the network sample $s=\{U_s, E_s\}$ obtained from the real world
network sampling design, we obtain a sequence of fast samples  
\[
\{S_1, S_2, S_3, ...,S_T \}
\] 
from the network data using a fast-sampling process 
similar to the original design.  $T$ is the number of iterations.

For unit $i \in U_s$, there is a sequence of indicator random variables:
\[
\{Z_{i1}, Z_{i2}, Z_{i3},..., Z_{iT} \}
\]
where $ Z_{it} = 1$ if $i\in S_t $ and  $Z_{it} = 0$ if $ i \notin
S_t$, for $t = 1, 2, ..., T$, the number of iterations of the sampling
process.  

The average 
\[
f_i = \frac{1}{T} \sum_{i = 1}^T Z_i
\]
is used as an estimate of the relative inclusion probability  of unit
$i$ in the similar design used to obtain the data from the real
world.  If the real-world network design is done without replacement,
then the fast-sampling process is also carried out without
replacement.  

In the first approach, each sample in the sequence proceeds from
selection of seeds to target sample size.  With this approach the
samples in the sequence $\{S_1, S_2, S_3, ...,S_T \}$ are independent
of each other.  

In the second approach, some seeds are selected in the beginning and
then each subsequent sample $S_t$ is selected dependent on the one
before it, $S_{t-1}$.  To get from $S_t$ to $S_{t-1}$ we
probabilistically trace links out from $S_t$, randomly drop some nodes
from $S_t$, and may with low probability select one or more new seeds.
Theoretical advantages of the second approach are first, that the
computation can be made very fast.  Second, the sampling process is
fast-mixing and once it reaches it's stationary distribution every
subsequent sample $S_t$ is in that distribution.  The stationary
distribution of the sequence of samples represents a balance between
the re-seeding distribution, which can be kept small with a low rate
of re-seeding, and the design tendencies arising from the link-tracing
and the without-replacement nature of the selections.  

In the independent sample approach, on the other hand, at each
iteration new seeds are selected. Usually the seed selection is from a distribution that is
different from the stationary distribution, so some number of waves of
link tracing is needed to get closer to the probabilities of the
original design.  Even when target sample
size is reached it may still be in a different distribution from the
original sample selection procedure because, with the necessarily
smaller sample size, the independent sample from the sample network
data is not able to go as many waves.  

Neither of the two computational implementation approaches can be
expected to be precisely unbiased, in the design-based sense, because
of the target sample size smaller than the actual sample size with the
without-replacement sampling, and because the generalized unequal
probability estimator, as a generalized ratio estimator, not being
precisely unbiased even when the true inclusion probabilities are
known.  The implementation of the independent, repeated sample
selection approach is straightforward based on the recruitment
protocol of the original design.  Because the fast sampling process
approach is less familiar, it is described in more detail here.

For the fast sample process $S_t$, instead of selecting
an entire an entire sample from seeds to target sample size, 
the approach is to set up a
sampling process.  For
the sampling process we start with seeds at iteration $t=0$ and at
each subsequent iteration a few
links may be traced from the links out from $S_t$  and a few nodes
removed from $S_t$ to obtain $S_{t+1}$.  Tracings and removals are
done with constant small probability and are independent, given
current sample size.  However, removals are only done if the sample
size of $S_t$ is above target, and then with probability calculated to
give an expected number of removals to go back to target.  

Specifically, in the examples we trace the links out from the current
sample $S_t$ independently, each with probability $p$.  Nodes are
removed from the sample independently with probability $q$.  The
removal probability $q$ is set adaptively to be $q_t = (n_t - n_{\rm
  target})/n_t$ if $n_t > n_{\rm target}$ and $q_t = 0$ otherwise, so
that sample size fluctuates around its target during iterations.
Sampling is without replacement in that a node in $S_t$ is not
reselected while it remains in fast sample, but it may be reselected
at any time after it is removed from the fast sample.

With this procedure, the sequence of fast samples $S_1, S_2, ...$
forms a Markov chain of sets, with the probability of set $S_t$
depending only on the previous set $S_{t-1}$.  
The
sampling process is fast mixing, with the sample
size fluctuating stochastically around the target.  

The inclusion-indicator average  $f_i$  converges in probability to
the inclusion probability for unit $i$ in the fast-sample design
process by the weak law of large numbers for Markov chains.  These in
turn approximate the inclusion probabilities of the real design to the
extent that the sampling process design is similar to the real design.
The effectiveness of this approximation must be evaluated with
simulations, as is done in the next section.

Alternatively, one can use a fast design that is with replacement.  An
advantage of this is that a target sample size for the fast design can
be used that equals the actual sample size used in obtaining the
data.  A disadvantage of using a with-replacement fast design is that
in many situations the real design by with the data are collected is
without-replacement.

If the fast-sample design is with replacement, let $M_t(i)$ be the number of times
node $i$ is selected at iteration $t$.  The quantity $g_i = (1/t)
\sum_{s=1}^t M_t(i)$, the average number of selections up to
iteration $t$, estimates the expected number of selections for node
$i$ under the with-replacement design at any given iteration $t$.

Fast sampling processes of these types are discussed in 
\cite{thompson2017adaptive} and \cite{thompson2015fast} for their
potential uses as measures of network exposure of a
node, or a measure of network centrality, or a predictive indicator of
regions of a network where an epidemic might next explode.
Calculation of the statistic $f_i$ for each unit in the network sample
can be used as an index of the network exposure of that unit.  A high value
of $f_i$ indicates the unit has high likelihood of being reached by a
network sample such as ours.  It will also have a relatively high
likelihood of being reached by a virus, such as HIV, that spreads on
the same type of links by a link-tracing process that is broadly
similar.  A given risk behavior will be more risky for a person with
high network exposure.  For a person in a less well connected part of
the network, the same behavior carries lower risk.  Since a purpose of
the surveys is to identify risk characteristics, an index of network
exposure measures another dimension of that risk, beyond the individual
behavior and health measures.  Here, however, are interested in their
usefulness for estimating population characteristics based on
link-tracing network sampling designs.

\section*{Estimators}

The network sampling designs considered here select units with unequal
probabilities.  With unequal probability sampling designs, sample
means and sample proportions do not provide unbiased estimates of
their corresponding population means and proportions.  

To estimate the mean of variable y with an unequal-probability
sampling design, the generalized unequal probability estimator has the
form 

\begin{equation}
\hat\mu_{\rm GUPE} = \frac{\sum_s (y_i/\pi_i)}{\sum_s (1/\pi_i)}
\end{equation}
where
$\pi_i$ is inclusion probability of unit $i$.

With the network sampling designs of interest here, the inclusion
probabilities $pi_i$ are not known and can not be calculated from the
sample data.  To circumvent this problem the Volz-Heckathorn Estimator
uses degree, or self-reported number of partners,  to approximate inclusion probability.

\begin{equation}
\hat\mu_{\rm VH} = \frac{\sum_s (y_i/d_i)}{\sum_s (1/d_i)}
\end{equation}
in which $d_i$ is the degree, the number of self-reported partners, of person $i$.

The rationale for this approximation is that if the sampling design is
a random walk with replacement, or several independent random walks with replacement and the population
network is connected, then the selection probabilities of the random
walk design will converge over time to be proportional to the $d_i$.
Here connected means that each node in the population can be reached
from any other node by some path, or chain of links, so that the
population network consists of only one connected component.  Biases
in this estimator result from the use of without-replacement sampling
in the real-world designs, the use of coupon numbers $k$ greater than
1 making the design different from a random walk, population networks
being not connected into a single component, or slow mixing due to
specifics of the
population network structure.    

The simple idea of the new estimators proposed here is to 
run a fast sampling process similar to the real design on the sample 
  network data. Then use the fast-sample inclusion frequencies to estimate 
the relative
  inclusion probabilities of the sample units.  The sample network
  data is the link part of the network sample.

The  simple estimator, with a non-replacement
sampling design, is 

\begin{equation}
\hat\mu = \frac{\sum_{i \in s} (y_i/f_i)}{\sum (1/f_i)}
\end{equation}
where there is no ambiguity we write $i \in s$ for the summation over
the nodes in the sample, rather than the more explicit $i \in U_s$,
and 
where
$f_i$ is \textit{inclusion frequency} of the fast-sampling process run
on the sample network data.

A simple variance estimator to go with the simple estimator is

\begin{equation}
\widehat{\textrm var}(\hat\mu) = \frac{1}{(\sum_{i \in s}
  1/f_i)^2}\sum_{i \in s} \frac{(y_i - \hat\mu)^2}{f_i^2}
\end{equation}

\medskip
\medskip

An approximate $1 - \alpha$ confidence interval is then calculated as 

\begin{equation}
\hat\mu \pm z \sqrt{\widehat{\textrm var}(\hat\mu)}
\end{equation}

with $z$ the $1-\alpha/2$ quantile from the standard Normal distribution.

The variance estimator is based on, and simplified from, the Taylor
series linear approximation theory for generalized unequal probability
estimator. 
Linearization leads to the estimator of the variance of the
generalized estimator 
\begin{equation}
\widehat{\textrm var}(\hat\mu_{\rm GUPE}) = \frac{1}{(\sum_{i \in s}
  1/\pi_i)^2}\sum_{i \in s} \sum_{j \in s} \check\Delta_{ij}\frac{(y_i
  - \hat\mu_{\rm GUPE})}{\pi_i}\frac{(y_j - \hat\mu_{\rm GUPE})}{\pi_j}
\end{equation}
where 
\[
\check\Delta_{ij} = \frac{\pi_{ij} - \pi_i\pi_j}{\pi_{ij}}
\]
where 
$\pi_{ij}$ is the joint inclusion probability for units $i$ and $j$.  
A good discussion of the approach is found in \cite{sarndal2003model},
with this variance estimator on p. 178 of that work.

\medskip

Consider an estimator of the variance using the full variance
expression with  the fast-sample
frequencies $f_i$ in place of the $\pi_i$ and, in place of the joint
inclusion probability $\pi_{ij}$, the frequency $f_{ij}$ of inclusion
of inclusion of both units $i$ and $j$ in the fast sampling process.
This would give 

\begin{equation}
\widehat{\textrm var}(\hat\mu) = \frac{1}{(\sum_{i \in s}
  1/f_i)^2}\sum_{i \in s} \sum_{j \in s} \hat\Delta_{ij}\frac{(y_i - \hat\mu)}{f_i}\frac{(y_j - \hat\mu)}{f_j}
\end{equation}
where 
\[
\hat\Delta_{ij} = \frac{f_{ij} - f_if_j}{f_{ij}}
\]

\medskip

The double sum in the variance estimate expression will have
$n(n-1)/2$ terms in which $i \ne j$.  The most influential of these
terms are the ones in which the joint frequency of inclusion $f_{ij}$
is relatively large.  Because of the link tracing in the fast sampling
process, sample unit pairs with a direct link between them will tend
occur together more frequently than those without a direct link.  An
estimator using only those pairs with known links between them in the
sample data would be 

\begin{equation}
\widehat{\textrm var}(\hat\mu) = \frac{1}{(\sum_{i \in s}
  1/f_i)^2} \left(\sum_{i \in U_s} (f_i - 1)\frac{(y_i -
    \hat\mu)^2}{f_i}  +  \sum_{(i,j) \in E_s} \hat\Delta_{ij}\frac{(y_i - \hat\mu)}{f_i}\frac{(y_j - \hat\mu)}{f_j}\right)
\end{equation}

where $E_s$ is the sample edge set.  That is, $E_s$ consists of the
known edges $(i,j)$ between pairs of units in the sample data.  In
general the size of the sample edge set $E_s$ will be much smaller
that the $n^2$ possible sample node pairings $(i, j)$, or the
$n(n-1)/2$ pairings with $i\ne j$, where $n$ is the sample size.  

A further simplification and approximation for estimating the variance
of the estimator is to use only the diagonal terms, that is, 

\begin{equation}
\widehat{\textrm var}(\hat\mu) = \frac{1}{(\sum_{i \in s}
  1/f_i)^2} \sum_{i \in s} (1 - f_i)\frac{(y_i -
    \hat\mu)^2}{f_i} 
\end{equation}

Dropping the coefficients $(1-f_i)$, each of which is less than or
equal to one, gives an estimate of variance that is larger, leading to
wider, more conservative confidence intervals.

If the real-world network sampling design and correspondingly the fast
sampling process are  with replacement, the estimator of $\mu$ is

\begin{equation}
\hat\mu =  \frac{\sum_{i \in s}
    (m_{i}y_i/g_i)}{\sum_{i \in s}
(m_{i}/g_i)}
\end{equation}
in which $m_i$ is the number of times unit i is selected in the real
design and $g_i$ is the average number of selection counts of unit
$i$ in the fast sampling process.

With a with-replacement fast design the corresponding variance
estimator is

\begin{equation}
\widehat{\textrm var}(\hat\mu) =   \frac{1}{(\sum_{i \in s}
  m_{i}/g_i)^2}\sum_{i \in s} \frac{m_{i}(y_i - \hat\mu)^2}{g_i^2}
\end{equation}

If $x_i$ is another variable, an estimator of the ratio
$R=\mu_y/\mu_x$ of the mean of
$y$ to the mean of $x$ is

\begin{equation}
\hat R = \frac{\sum_{i \in s} y_i/f_i}{\sum_{i\in s} x_i/f_i}
\end{equation}

with simple variance estimator

\begin{equation}
\widehat{\textrm var}(\hat R) = \frac{1}{(\sum_{i \in s}
  x_i/f_i)^2}\sum_{i \in s} \frac{(y_i - x_i\hat\mu)^2}{f_i^2}
\end{equation}

The source C code of the algorithm for the fast sampling processes and
calculation of the inclusion frequencies $f_i$ on which the estimators
are based can be found at stevenkthompson/simple on
https://github.com.

\section*{Simulations}

For an empirical population with which to evaluate the effectiveness
of the proposed inference method we use the network data from the
Colorado Springs Study on the heterosexual spread of HIV.  Called the
Project 90 study, it was carried out to delineate the behaviors and
network characteristics of a high risk population including sex
workers, clients of sex workers, drug users and associated people.
The study was carried out by members of the El Paso County Health
Department and was supported by the Centers for Disease Control and
Prevention (CDC) \cite{potterat1999network}.  In this study, every
effort was made to trace the sexual, drug, and social relationships of
the members of this at-risk population, identifying names and
characteristics of each partner and later destroying the data on names
or unique identifiers while retaining the anonymized network
structure.  Since this effort was far more thorough than is usual the
case in network studies, this network data set is highly suitable as a
simulation population from which to sample and evaluate network
sampling and inference methods.  For studies of hidden and
hard-to-reach human populations, this is the most relevant of
available network data sets for simulations evaluating methodologies.

We use here the complete data set placed by Matt Salganik in the
archive maintained by the Princeton Office of Population Research 
\break (https://opr.princeton.edu/archive/p90/). 
The data set has N = 5492 people and  L = 21,644 links between those
people.  The links represent social, sexual, and drug-related
relationships and are symmetric.  The data set includes 13 attribute
variables on each person, including gender (male = 0, female = 1), sex
work, client of sex worker, unemployed and other variables, some
risk-related and some not, described in the data
archive.  The links are not identified by type, so the presence of a
link between two people indicates there is at least one of the
relationship types social, sexual, or drug related.  

The data network of 5492 nodes and 21,644 has 108 separate components,
the largest of which contains 4430 nodes, followed by components of
sizes 50, 42, 26, and so on down to components of size 2.  

For the simulation I used the entire Project 90 archive data set with
its 108 components as more realistically representing other real
populations of interest than would using using solely the largest
component, as has sometimes been done in other methodology studies.
All missing item responses have been arbitrarily changed to zero, so
that the number of nodes and sample sizes for each of the variables
are the same and so comparisons of estimation methods are not influenced by
different sample sizes for different variables or by choices in how to
handle missing values.

Option and rate values uses in the simulations were as follow.  For
the original without-replacement the coupon limit was 3 for RDS and
RDS+ and 25 for SB and SB+.  Less that 4 percent of people in the
population have degree more than 25, so that with the snowball designs
the number of coupons would equal their number of partners for more
than 96 percent of the population.  The target number of seeds was 240
and the target sample size was 1200.  Seeds were selected by Bernoulli
random selections with the desired expected sample size; coupons had
an expiration date 28 days from date of issue; each day before sample
size was reached; there was Bernoulli probability 0.00001 of selecting
re-seeds, but this was hardly ever done; from the units outside the
sample.  The Bernoulli probability each day of tracing any link out of
the sample where the coupon had not expired was 0.004.  On average, 19
percent of the links within the sample were traced by coupon
redemptions.

For the without-replacement fast sampling process the target sample
size was 400; the Bernoulli link tracing probability on links out was
0.5 and the Bernoulli probability of node removals from the sample
was also 0.5; re-seed Bernoulli probability for the nodes outside the
sample was 0.1.  The fast sampling process was the same, free to
follow links and unrestrained by coupons, for each of the four
original designs.

In each simulation run the new estimator is calculated along with the
VH estimator and the sample mean.  Like the new estimator, the VH
works the same for all types of variables, whether binary or more
general numeric, and does not  not require an estimate of population size.  
Because of the relatively small sampling fraction .22 in the
simulations, the SS estimator here would be expected to be similar to
the VH estimator, if the estimate of N for the SS estimator was accurate
or biased upward.  The SH estimator would be identical to the VH
estimator for estimating mean degree.  The HCG would be similar to the
VH for estimating mean degree, because of the small sample fraction.

Separate simulations were done for each of the four sampling designs
RDS, RDS+, SB, and SB+.  
Simulations for each of the four designs were carried out with the
following steps.

\begin{enumerate}

\item  From the population of N = 5492 people and L = 21,644 links, select
a target sample of n = 1200 people,
using RDS, RDS+, SB, or SB+.    To start each sample, a target  of 240 seeds are selected at
random.  A small number of additional seeds may be selected in the
process of reaching the target sample size of 1200.

\item Calculate the $f_i$ with fast samples $S_1, S_2, ...,
  S_T$.  The number of iterations $T$ used was 10,000.  The target
  sample size for each $S_t$ in the sequence was 400, or one-third of
  the sample size $n = 1200$.  
 Estimate means of degree and 13 other variables, using the new
 estimator $\hat \mu$, the Volz-Heckathorn estimator $\hat\mu_{VH}$
 and the sample mean or proportion.

\end{enumerate}

Steps 1 and 2 were repeated 1000 times for each of the four designs.

\medskip

Using the 1000 values for each estimators, together with the known
population values, the mean square error (MSE), Bias, and Confidence
Interval Coverage were calculated.  

Confidence intervals were calculated for the new estimator using
the simple variance estimator described in the previous section.
Confidence interval coverage for each estimate was calculated as the
proportion of intervals, out of 1000, that covered the true mean of
the variable.

\section*{Results}

The results of the simulations are  summarized in the figures.  The numerical results on which the
figures are based are in the tables of \nameref{S1_Tables}.  Figures 1
and 2 show the results for estimating mean degree.  Fig 1 plots the
MSE for estimating mean degree with the new estimator (vertical axis)
against the MSE of the VH estimator for each of the four designs RDS,
RDS+, SB, and SB+.  For a point below the diagonal line the MSE is
lower with the new estimator.  Each point here is far below the line.
The relative efficiencies (MSE(VH)/MSE(Simple)) corresponding to the
four points and given in Tables 1-4 of \nameref{S1_Tables}, are 29
for RDS, 54 for RDS+, 26 for SB and 80 for SB+.The relative efficiencies (MSE(VH)/MSE(Simple)) corresponding to the
four points and given in Tables 1-4 of \nameref{S1_Tables}, are 29
for RDS, 54 for RDS+, 26 for SB and 80 for SB+.  This means for
example that with a standard RDS study and using the same data, the
new estimator has one-twenty-ninth the MSE as does the VH estimator.
With each of the four designs, the MSE is large for the VH estimator
and is low for the new estimator.  The lower mean square errors of the
new estimator are achieved by eliminating most of the bias.  

\bigskip
\noindent\textbf{Fig 1.}  For estimating mean degree, MSE of the new
  estimator (vertical axis) is plotted against MSE of the VH
  estimator.  Each point represents a design.  The new estimator has
  much lower MSE for each of the designs.  
\bigskip

A measure of how much is gained by identifying the
within-sample links beyond those used in recruitment is obtained by
comparing the relative efficiency for estimating mean degree using the
simple estimator for each of the ``plus'' designs with the
corresponding standard designs in Fig 1.  Using the simple estimator in each
case, the efficiency gains are MSE(RDS)/MSE(RDS+) = 1.83 and
MSE(SB)/MSE(SB+) = 3.00 (using the MSE values from Tables 1-4  of  \nameref{S1_Tables}).

Figure 2 shows the same relationships but for the bias of the
estimator of mean degree instead of MSE.
Each point again represents a design and the absolute bias with the
new estimator is on the vertical axis while the absolute bias for the
VH estimator  with the four designs.
designs.  For instance, with RDS, the bias in estimating mean degree
with the new estimator is .32 while the bias with VH is -2.46 (Table
1  of  \nameref{S1_Tables}).  For each of the four designs the bias of the VH estimator is
large and the bias of the new estimator.  Further, the bias of the VH
estimator is worse with the SB and SB+ designs using 25 coupons than
it is with the RDS and RDS+ designs using 3 coupons.  That is to be
expected since the VH estimator is based on the approximating
assumption of a random walk design, and the more coupons handed out
the more branching in recruitment, and the designs get even farther
from the assumed random walk. 

\bigskip
\noindent\textbf{Fig 2.}  Bias in estimating mean degree. Bias of the new
  estimator (vertical axis) is plotted against bias of the VH
  estimator.  Each point represents a different design.  The new
  estimator has much less bias for each of the designs.  
\bigskip

%Table 1

Also we see in Fig 2 that the new estimator is able to take advantage
of the extra link information of the enhanced (+) designs to further
reduce bias, while the VH estimator is not able to benefit from the
extra sample network information of the enhanced designs.

% Table 2-4

For estimating the population proportion of each of the 13 attribute
variables, the MSE comparisons are given in the four Figures 3-6.
Here each plot is for just one of the designs RDS, RDS+, SB, and SB+.  Each
point represents a different binary variable variable, as identified
in the legend.  Once again,
a point below the diagonal indicates the MSE is lower with the new
estimator compared to the VH estimator.  The pattern in each plot is that the MSE is lower with the
new estimator than with the VH estimator, except for some of the
points near the origin where the MSE is very small with both
estimators. 

\bigskip
\noindent\textbf{Fig 3.}   MSE with design RDS for estimating
population proportion for each of the 13 attribute variables.   
\bigskip

\bigskip
\noindent\textbf{Fig 4.}   MSE with design RDS+ for estimating
population proportion for each of the 13 attribute variables.   
\bigskip

\bigskip
\noindent\textbf{Fig 5.}   MSE with design SB for estimating
population proportion for each of the 13 attribute variables.   
\bigskip

\bigskip
\noindent\textbf{Fig 6.}   MSE with design SB+ for estimating
population proportion for each of the 13 attribute variables.   
\bigskip

 A similar pattern for bias is shown in the four Figures
7-10.  While neither estimator has uniformly lower bias for all
variables, the new estimator has the lower bias except for cases where
the bias is small with either estimator.  

\bigskip
\noindent\textbf{Fig 7.}    Bias with design RDS for estimating
population proportion for each of the 13 attribute variables. 
\bigskip

\bigskip
\noindent\textbf{Fig 8.}    Bias with design RDS+ for estimating
population proportion for each of the 13 attribute variables. 
\bigskip

\bigskip
\noindent\textbf{Fig 9.}    Bias with design SB for estimating
population proportion for each of the 13 attribute variables. 
\bigskip

\bigskip
\noindent\textbf{Fig 10.}    Bias with design SB+ for estimating
population proportion for each of the 13 attribute variables. 
\bigskip

The exact MSE and Bias numeric results are shown in Tables 1-4  of  \nameref{S1_Tables}.  The
last two columns on the right in those tables show relative
efficiency of the new estimator compared to VH or sample mean, and
relative absolute bias.  The column ``eff'' gives in the VH section of
the table gives MSE(VH)/MSE(Simple) for each of the variables, and
``rbias'' gives $|$Bias(VH)$|$/$|$Bias(Simple)$|$.  Values above 1.0
are favorable to the new estimator and values below 0.0 are favorable
to VH, and show the details of the points clustered near the diagonal
line and origin in the plots.  The unweighted sample means, which have
higher biases in most but not all cases, are included in the tables
for reference.

Confidence interval widths and coverage probabilities were calculated
only for the proposed estimator and are given in Tables 5-8 of
\nameref{S1_Tables}, with each table for one of the four designs. The
target coverage is .95.  ``AV $\widehat{SD}$'' is the square root of
the average value of $\widehat{{\rm var}}(\hat\mu)$ over the 1000
runs.  ``AV width/2'' is the average half-width of the confidence
interval.  ``Coverage'' in the tables is the coverage probability, the
proportion of the 1000 simulation runs in which the confidence
interval covers the true value.

% Table 7
% sb
% print(xtable(data.frame(name=data25a$name[1:14], actual=data25a$actual[1:14], halfwidth=ci25rdata$half.width, coverage=ci25rdata$coverage), caption='SB: Confidence Interval Coverage', label='SBcoverage'), include.rownames = FALSE)
% latex table generated in R 3.4.4 by xtable 1.8-2 package
% Sun Sep  2 11:42:59 2018

% Table 8
% sbplus
% print(xtable(data.frame(name=data25a$name[1:14], actual=data25a$actual[1:14], halfwidth=ci25adata$half.width, coverage=ci25adata$coverage), caption='SB+: Confidence Interval Coverage', label='SBPcoverage'), include.rownames = FALSE)
% latex table generated in R 3.4.4 by xtable 1.8-2 package
% Sat Sep  1 17:54:29 2018

Most of the coverage probabilities and in the 80s and 90s percents.
Where coverage is lower than that, it is in most cases because the
estimator has a bias similar in magnitude to it's standard deviation,
so that a symmetric confidence interval based on an estimate of that
standard deviation will tend to be off-center.  For this reason this
same confidence interval method would not be recommended for the VH
estimator with it's tendency to have larger biases.

Although the simulations here compare the new estimator directly only
to the VH estimator and the sample mean, the estimate of mean degree
with the SS estimator is the same as with the VH estimator, so the SS
estimate of mean degree will have the same bias and MSE as the VH
estimate.  With the sampling fraction $n/N = 0.22$ here, the SS
and HCG estimates of mean degree should also be approximately like the
VH estimator.  

The \nameref{S1_Tables} also document the historical advance of the
estimators such as VH and SH over what went before them.  Earlier
link-tracing studies of hidden populations by field ethnographers,
anthropologists, epidemiologists had no alternatives to reporting the
results of their findings as sample means and proportions
(\cite{heckathorn1997respondent} \cite{thompson2002adaptive} include
reviews of earlier studies).  From the tables we can see for example
for estimating population mean degree, the efficiency gain of the VH,
and for SH with is the same for mean degree, is for RDS 41.432166 /
6.092061 = 6.80, for RDS+ 41.272397 / 6.089313 = 6.78, for SB
40.406463 / 7.272708 = 5.56, and for SB+ 40.369395 / 7.306935 = 5.52.

Comparing the new estimators to the sample means and proportions the
relative efficiencies can be read  directly from the ``eff'' column of
the  \nameref{S1_Tables}.  The efficiency
gain of the new estimator over the sample mean for estimating
population mean degree is for RDS 198.54, for RDS+ 362.80, for SB
146.54, and for SB+ 439.49.  The new estimators are made possible by
better understanding of networks and of network sampling designs,
faster computers, and computational data structures and algorithms
which were not available two decades ago.

\section*{Discussion and conclusions}

The simple estimators introduced here for network sampling use the
sample network data and a fast sampling process similar to the actual
sampling design used to collect the data.  Inclusion frequencies of
the fast sampling process estimate the relative inclusion
probabilities of the actual design.  These estimated inclusion
probabilities are in turn used to form the sampling weights for
estimates of population means and proportions.  

The new estimators move beyond the approximating Markov chain
assumptions about the recruitment of people into the sample or about
the transitions in the recruitments different values of binary
variables.  The new estimators do not assume that the population
network has only a single connected component.  The discrepancy
between such assumptions and reality has led to biases and high mean
square errors in the estimators that depend on them.

The availability of these new estimators should free investigators to
use different types of network sampling designs as they wish.  For
instance, the new estimators work as well with snowball designs that
are virtually unlimited by imposed coupon limits as they do for
designs that limit recruitment coupons to few.  The practice of
restricting recruitment coupons to small numbers such as two or three
might make sense for some studies, in order to reach farther into a
hidden population as opposed to exploring key areas of the population
more thoroughly.  The use of small coupon numbers was motivated
originally by the hope that it would make the sampling closer to a
random walk so that the estimation biases would not be too large.
With the new estimators investigators can choose whatever coupon
limits make sense for the study.  Also for studies of at-risk key
populations it may make sense to follow risk related links, such as
drug injection related links or sexual links.  With the new
estimators, it is not necessary to follow more general links such as
friendships in the hope that they connect the entire population into a
single component.

The new estimators work as well for quantitative or continuous
variables as for binary variables.  The assumption about Markov
transitions between states of a binary variable have led some
estimation methods to be developed only for estimating the means of
binary variables.  In modern network surveys of hidden populations
some of the most important variables are quantitative, such as number of
sessions of sex in a period or frequency of drug injection activity,
as well as number of partners.  These quantitative variables are also
the most likely to be related to the sampling inclusion probabilities
and so benefit the most from better estimates of those inclusion
probabilities.  

Existing data from network sampling studies can be reanalyzed using
the new methods.  For new studies investigators could consider using
the enhanced data collection methods that obtain more information
about within-sample links, that produce better estimates with the same
number of people in the sample.

\section*{Supporting information}

% Include only the SI item label in the paragraph heading. Use the \nameref{label} command to cite SI items in the text.
\paragraph*{S1 Tables.  }
\label{S1_Tables}
{\bf Tables for Simple estimators for network
  sampling.} 

\bigskip

Tables 1-4 correspond  to the four network
  sampling designs RDS, RDS+, SB, and SB+ respectively.  

\bigskip

Tables 5-8 give confidence
  interval coverages and widths for the same four designs.

\section*{Acknowledgments}

This research was supported by Natural Science and Engineering
Research Council of Canada (NSERC) Discovery grant RGPIN327306.  I would like
to thank John Potterat and Steve Muth for making the Project 90 data
available and Matt Salganik for making part of it conveniently
available on the Princeton Population Center archiv for the research
community. I would like to express appreciation for the participants
in that study who shared their personal information with the
researchers so that it could be made available in anonymized form to
the research community and contribute to a solution to HIV and
addiction epidemics and to basic understanding of social networks.

\nolinenumbers

%Table 5
%rds
%print(xtable(data.frame(name=data25a$name[1:14], actual=data25a$actual[1:14], halfwidth=ci3rdata$half.width, coverage=ci3rdata$coverage), caption='RDS: Confidence Interval Coverage', label='RDScoverage'), include.rownames = FALSE)
% latex table generated in R 3.4.4 by xtable 1.8-2 package
% Sat Sep  1 17:33:05 2018

% Table 6
% rdsplus
% print(xtable(data.frame(name=data25a$name[1:14], actual=data25a$actual[1:14], halfwidth=ci3adata$half.width, coverage=ci3adata$coverage), caption='RDS+: Confidence Interval Coverage', label='RDSPcoverage'), include.rownames = FALSE)
% latex table generated in R 3.4.4 by xtable 1.8-2 package
% Sun Sep  2 11:43:00 2018

%\section*{References}
% The bibtex filename
% uncomment the following line when new references are added to
%   fastrefs.bib, then copy the resulting .bbl file to below:
\bibliography{fastrefs}

\pagebreak

\begin{flushleft}
{\Large
\textbf\newline{\textbf{S1.  Tables for Simple estimators for network
  sampling.} }

\bigskip

Tables 1-4 correspond  to the four network
  sampling designs RDS, RDS+, SB, and SB+ respectively.  

\bigskip

Tables 5-8 give confidence
  interval coverages and widths for the same four designs.  }
\end{flushleft}

\bigskip

%> print(xtable(data3r, caption='RDS Table', label='rdstable', digits=c(2,2,2,2,6,6,6,2,2)), include.rownames = F)
% latex table generated in R 3.4.4 by xtable 1.8-2 package
% Wed Oct  3 15:22:01 2018
\begin{table}[ht]
\centering
\caption{RDS Design} 
\begin{tabular}{lrrrrrrr}
  \hline
SIMPLE & actual & E.est & bias & sd & mse & eff & rbias \\ 
  \hline
degree & 7.88 & 8.20 & 0.321723 & 0.324316 & 0.208686 & 1.00 & 1.00 \\ 
  nonwhite & 0.24 & 0.25 & 0.010975 & 0.021323 & 0.000575 & 1.00 & 1.00 \\ 
  female & 0.43 & 0.43 & 0.001933 & 0.021844 & 0.000481 & 1.00 & 1.00 \\ 
  worker & 0.05 & 0.06 & 0.003294 & 0.009343 & 0.000098 & 1.00 & 1.00 \\ 
  procurer & 0.02 & 0.02 & 0.001547 & 0.004637 & 0.000024 & 1.00 & 1.00 \\ 
  client & 0.09 & 0.09 & 0.000607 & 0.014468 & 0.000210 & 1.00 & 1.00 \\ 
  dealer & 0.06 & 0.07 & 0.006345 & 0.010341 & 0.000147 & 1.00 & 1.00 \\ 
  cook & 0.01 & 0.01 & 0.000187 & 0.004091 & 0.000017 & 1.00 & 1.00 \\ 
  thief & 0.02 & 0.02 & 0.001593 & 0.006493 & 0.000045 & 1.00 & 1.00 \\ 
  retired & 0.03 & 0.03 & 0.000251 & 0.007735 & 0.000060 & 1.00 & 1.00 \\ 
  homemakr & 0.06 & 0.06 & -0.000003 & 0.010903 & 0.000119 & 1.00 & 1.00 \\ 
  disabled & 0.04 & 0.04 & 0.000857 & 0.008706 & 0.000077 & 1.00 & 1.00 \\ 
  unemploy & 0.16 & 0.17 & 0.006356 & 0.015811 & 0.000290 & 1.00 & 1.00 \\ 
  homeless & 0.01 & 0.01 & 0.000491 & 0.005062 & 0.000026 & 1.00 & 1.00 \\ 
    \hline
VH & actual & E.est & bias & sd & mse & eff & rbias \\ 
  \hline
  degree & 7.88 & 5.42 & -2.458292 & 0.221049 & 6.092061 & 29.19 & 7.64 \\ 
  nonwhite & 0.24 & 0.26 & 0.020619 & 0.020615 & 0.000850 & 1.48 & 1.88 \\ 
  female & 0.43 & 0.41 & -0.024778 & 0.021961 & 0.001096 & 2.28 & 12.82 \\ 
  worker & 0.05 & 0.05 & -0.004860 & 0.010110 & 0.000126 & 1.28 & 1.48 \\ 
  procurer & 0.02 & 0.01 & -0.003472 & 0.003454 & 0.000024 & 1.00 & 2.24 \\ 
  client & 0.09 & 0.13 & 0.038377 & 0.017908 & 0.001793 & 8.55 & 63.20 \\ 
  dealer & 0.06 & 0.07 & 0.001294 & 0.010499 & 0.000112 & 0.76 & 0.20 \\ 
  cook & 0.01 & 0.01 & -0.001309 & 0.002985 & 0.000011 & 0.63 & 7.00 \\ 
  thief & 0.02 & 0.02 & -0.000879 & 0.006085 & 0.000038 & 0.85 & 0.55 \\ 
  retired & 0.03 & 0.03 & 0.002616 & 0.008131 & 0.000073 & 1.22 & 10.41 \\ 
  homemakr & 0.06 & 0.05 & -0.008380 & 0.008862 & 0.000149 & 1.25 & 2627.83 \\ 
  disabled & 0.04 & 0.04 & -0.005227 & 0.007436 & 0.000083 & 1.08 & 6.10 \\ 
  unemploy & 0.16 & 0.13 & -0.027922 & 0.013437 & 0.000960 & 3.31 & 4.39 \\ 
  homeless & 0.01 & 0.01 & -0.000674 & 0.004199 & 0.000018 & 0.70 & 1.37 \\ 
  \hline
$\bar y$ & actual & E.est & bias & sd & mse & eff & rbias \\ 
  \hline
  degree & 7.88 & 14.31 & 6.432272 & 0.240921 & 41.432166 & 198.54 & 19.99 \\ 
  nonwhite & 0.24 & 0.28 & 0.039685 & 0.016599 & 0.001850 & 3.22 & 3.62 \\ 
  female & 0.43 & 0.46 & 0.032543 & 0.011160 & 0.001184 & 2.46 & 16.83 \\ 
  worker & 0.05 & 0.09 & 0.040418 & 0.006369 & 0.001674 & 17.06 & 12.27 \\ 
  procurer & 0.02 & 0.03 & 0.015857 & 0.003242 & 0.000262 & 10.96 & 10.25 \\ 
  client & 0.09 & 0.07 & -0.015363 & 0.006984 & 0.000285 & 1.36 & 25.30 \\ 
  dealer & 0.06 & 0.12 & 0.054326 & 0.007186 & 0.003003 & 20.40 & 8.56 \\ 
  cook & 0.01 & 0.01 & 0.001482 & 0.002368 & 0.000008 & 0.47 & 7.93 \\ 
  thief & 0.02 & 0.04 & 0.014693 & 0.004091 & 0.000233 & 5.20 & 9.22 \\ 
  retired & 0.03 & 0.03 & -0.000726 & 0.004055 & 0.000017 & 0.28 & 2.89 \\ 
  homemakr & 0.06 & 0.07 & 0.007519 & 0.006176 & 0.000095 & 0.80 & 2357.91 \\ 
  disabled & 0.04 & 0.06 & 0.014319 & 0.005355 & 0.000234 & 3.05 & 16.70 \\ 
  unemploy & 0.16 & 0.25 & 0.090330 & 0.010243 & 0.008264 & 28.46 & 14.21 \\ 
  homeless & 0.01 & 0.02 & 0.003988 & 0.002808 & 0.000024 & 0.92 & 8.12 \\ 
   \hline
\end{tabular}
\label{rdstable}
\end{table}

%> print(xtable(data3a, caption='RDS+ Design', label='rdsplustable', digits=c(2,2,2,2,6,6,6,2,2)), include.rownames = F)
% latex table generated in R 3.4.4 by xtable 1.8-2 package
% Wed Oct  3 15:49:28 2018
\begin{table}[ht]
\centering
\caption{RDS+ Design} 
\begin{tabular}{lrrrrrrr}
  \hline
SIMPLE & actual & E.est & bias & sd & mse & eff & rbias \\ 
  \hline
degree & 7.88 & 7.82 & -0.064183 & 0.331123 & 0.113762 & 1.00 & 1.00 \\ 
  nonwhite & 0.24 & 0.22 & -0.016216 & 0.023923 & 0.000835 & 1.00 & 1.00 \\ 
  female & 0.43 & 0.44 & 0.011235 & 0.027851 & 0.000902 & 1.00 & 1.00 \\ 
  worker & 0.05 & 0.05 & 0.002313 & 0.011178 & 0.000130 & 1.00 & 1.00 \\ 
  procurer & 0.02 & 0.01 & -0.000711 & 0.004308 & 0.000019 & 1.00 & 1.00 \\ 
  client & 0.09 & 0.07 & -0.017189 & 0.015128 & 0.000524 & 1.00 & 1.00 \\ 
  dealer & 0.06 & 0.06 & -0.006107 & 0.009508 & 0.000128 & 1.00 & 1.00 \\ 
  cook & 0.01 & 0.01 & -0.000597 & 0.004478 & 0.000020 & 1.00 & 1.00 \\ 
  thief & 0.02 & 0.02 & -0.002592 & 0.005953 & 0.000042 & 1.00 & 1.00 \\ 
  retired & 0.03 & 0.03 & -0.002589 & 0.008533 & 0.000080 & 1.00 & 1.00 \\ 
  homemakr & 0.06 & 0.06 & 0.000696 & 0.013245 & 0.000176 & 1.00 & 1.00 \\ 
  disabled & 0.04 & 0.04 & -0.001669 & 0.009565 & 0.000094 & 1.00 & 1.00 \\ 
  unemploy & 0.16 & 0.16 & -0.001582 & 0.018396 & 0.000341 & 1.00 & 1.00 \\ 
  homeless & 0.01 & 0.01 & -0.000682 & 0.005280 & 0.000028 & 1.00 & 1.00 \\ 
   \hline
VH & actual & E.est & bias & sd & mse & eff & rbias \\ 
  \hline
  degree & 7.88 & 5.42 & -2.457134 & 0.227604 & 6.089313 & 53.53 & 38.28 \\ 
  nonwhite & 0.24 & 0.26 & 0.020821 & 0.021395 & 0.000891 & 1.07 & 1.28 \\ 
  female & 0.43 & 0.41 & -0.024457 & 0.021244 & 0.001049 & 1.16 & 2.18 \\ 
  worker & 0.05 & 0.05 & -0.004943 & 0.009720 & 0.000119 & 0.91 & 2.14 \\ 
  procurer & 0.02 & 0.01 & -0.003419 & 0.003376 & 0.000023 & 1.21 & 4.81 \\ 
  client & 0.09 & 0.13 & 0.037166 & 0.018657 & 0.001729 & 3.30 & 2.16 \\ 
  dealer & 0.06 & 0.06 & 0.000927 & 0.010374 & 0.000108 & 0.85 & 0.15 \\ 
  cook & 0.01 & 0.01 & -0.001510 & 0.002819 & 0.000010 & 0.50 & 2.53 \\ 
  thief & 0.02 & 0.02 & -0.000778 & 0.005826 & 0.000035 & 0.82 & 0.30 \\ 
  retired & 0.03 & 0.03 & 0.002589 & 0.008051 & 0.000072 & 0.90 & 1.00 \\ 
  homemakr & 0.06 & 0.05 & -0.008324 & 0.008794 & 0.000147 & 0.83 & 11.95 \\ 
  disabled & 0.04 & 0.04 & -0.005882 & 0.007681 & 0.000094 & 0.99 & 3.52 \\ 
  unemploy & 0.16 & 0.13 & -0.029356 & 0.013133 & 0.001034 & 3.03 & 18.56 \\ 
  homeless & 0.01 & 0.01 & -0.000674 & 0.004227 & 0.000018 & 0.65 & 0.99 \\ 
  \hline
$\bar y$ & actual & E.est & bias & sd & mse & eff & rbias \\ 
  \hline
  degree & 7.88 & 14.30 & 6.419863 & 0.240330 & 41.272397 & 362.80 & 100.02 \\ 
  nonwhite & 0.24 & 0.28 & 0.039531 & 0.017219 & 0.001859 & 2.23 & 2.44 \\ 
  female & 0.43 & 0.47 & 0.032806 & 0.011357 & 0.001205 & 1.34 & 2.92 \\ 
  worker & 0.05 & 0.09 & 0.040330 & 0.005867 & 0.001661 & 12.75 & 17.44 \\ 
  procurer & 0.02 & 0.03 & 0.016056 & 0.003326 & 0.000269 & 14.10 & 22.57 \\ 
  client & 0.09 & 0.07 & -0.015742 & 0.007279 & 0.000301 & 0.57 & 0.92 \\ 
  dealer & 0.06 & 0.12 & 0.054282 & 0.006950 & 0.002995 & 23.45 & 8.89 \\ 
  cook & 0.01 & 0.01 & 0.001442 & 0.002335 & 0.000008 & 0.37 & 2.42 \\ 
  thief & 0.02 & 0.04 & 0.014605 & 0.004005 & 0.000229 & 5.44 & 5.63 \\ 
  retired & 0.03 & 0.03 & -0.000696 & 0.003973 & 0.000016 & 0.20 & 0.27 \\ 
  homemakr & 0.06 & 0.07 & 0.007271 & 0.006181 & 0.000091 & 0.52 & 10.44 \\ 
  disabled & 0.04 & 0.06 & 0.014378 & 0.005299 & 0.000235 & 2.49 & 8.61 \\ 
  unemploy & 0.16 & 0.25 & 0.089440 & 0.010143 & 0.008102 & 23.77 & 56.54 \\ 
  homeless & 0.01 & 0.02 & 0.003962 & 0.002762 & 0.000023 & 0.82 & 5.81 \\ 
   \hline
\end{tabular}
\label{rdsplustable}
\end{table}

%> print(xtable(data25r, caption='SB Design', label='sbtable', digits=c(2,2,2,2,6,6,6,2,2)), include.rownames = F)
% latex table generated in R 3.4.4 by xtable 1.8-2 package
% Wed Oct  3 15:51:08 2018
\begin{table}[ht]
\centering
\caption{SB Design} 
\begin{tabular}{lrrrrrrr}
  \hline
SIMPLE & actual & E.est & bias & sd & mse & eff & rbias \\ 
  \hline
degree & 7.88 & 8.32 & 0.441944 & 0.283595 & 0.275741 & 1.00 & 1.00 \\ 
  nonwhite & 0.24 & 0.26 & 0.016708 & 0.021857 & 0.000757 & 1.00 & 1.00 \\ 
  female & 0.43 & 0.43 & 0.000794 & 0.022492 & 0.000507 & 1.00 & 1.00 \\ 
  worker & 0.05 & 0.06 & 0.005829 & 0.009198 & 0.000119 & 1.00 & 1.00 \\ 
  procurer & 0.02 & 0.02 & 0.003119 & 0.004730 & 0.000032 & 1.00 & 1.00 \\ 
  client & 0.09 & 0.09 & 0.004746 & 0.013939 & 0.000217 & 1.00 & 1.00 \\ 
  dealer & 0.06 & 0.07 & 0.010055 & 0.010421 & 0.000210 & 1.00 & 1.00 \\ 
  cook & 0.01 & 0.01 & 0.000193 & 0.003848 & 0.000015 & 1.00 & 1.00 \\ 
  thief & 0.02 & 0.03 & 0.003107 & 0.006341 & 0.000050 & 1.00 & 1.00 \\ 
  retired & 0.03 & 0.03 & 0.000531 & 0.008084 & 0.000066 & 1.00 & 1.00 \\ 
  homemakr & 0.06 & 0.06 & -0.001243 & 0.010849 & 0.000119 & 1.00 & 1.00 \\ 
  disabled & 0.04 & 0.04 & 0.001952 & 0.008947 & 0.000084 & 1.00 & 1.00 \\ 
  unemploy & 0.16 & 0.17 & 0.010233 & 0.015945 & 0.000359 & 1.00 & 1.00 \\ 
  homeless & 0.01 & 0.01 & 0.000955 & 0.005008 & 0.000026 & 1.00 & 1.00 \\ 
  \hline
VH & actual & E.est & bias & sd & mse & eff & rbias \\ 
  \hline
  degree & 7.88 & 5.19 & -2.689333 & 0.200484 & 7.272708 & 26.38 & 6.09 \\ 
  nonwhite & 0.24 & 0.27 & 0.031928 & 0.021830 & 0.001496 & 1.98 & 1.91 \\ 
  female & 0.43 & 0.39 & -0.040046 & 0.021358 & 0.002060 & 4.07 & 50.42 \\ 
  worker & 0.05 & 0.05 & -0.002207 & 0.008953 & 0.000085 & 0.72 & 0.38 \\ 
  procurer & 0.02 & 0.01 & -0.001881 & 0.003648 & 0.000017 & 0.52 & 0.60 \\ 
  client & 0.09 & 0.15 & 0.064678 & 0.019204 & 0.004552 & 21.00 & 13.63 \\ 
  dealer & 0.06 & 0.07 & 0.008575 & 0.010947 & 0.000193 & 0.92 & 0.85 \\ 
  cook & 0.01 & 0.01 & -0.001647 & 0.002800 & 0.000011 & 0.71 & 8.54 \\ 
  thief & 0.02 & 0.02 & 0.002400 & 0.006449 & 0.000047 & 0.95 & 0.77 \\ 
  retired & 0.03 & 0.03 & 0.005040 & 0.008598 & 0.000099 & 1.51 & 9.48 \\ 
  homemakr & 0.06 & 0.05 & -0.012866 & 0.008499 & 0.000238 & 1.99 & 10.35 \\ 
  disabled & 0.04 & 0.04 & -0.006002 & 0.007144 & 0.000087 & 1.04 & 3.07 \\ 
  unemploy & 0.16 & 0.13 & -0.030449 & 0.012518 & 0.001084 & 3.02 & 2.98 \\ 
  homeless & 0.01 & 0.01 & -0.000407 & 0.004111 & 0.000017 & 0.66 & 0.43 \\ 
  \hline
$\bar y$ & actual & E.est & bias & sd & mse & eff & rbias \\ 
  \hline
  degree & 7.88 & 14.24 & 6.353473 & 0.199600 & 40.406463 & 146.54 & 14.38 \\ 
  nonwhite & 0.24 & 0.30 & 0.057156 & 0.017098 & 0.003559 & 4.70 & 3.42 \\ 
  female & 0.43 & 0.46 & 0.025163 & 0.011382 & 0.000763 & 1.51 & 31.68 \\ 
  worker & 0.05 & 0.10 & 0.047583 & 0.005640 & 0.002296 & 19.36 & 8.16 \\ 
  procurer & 0.02 & 0.03 & 0.019414 & 0.003096 & 0.000386 & 12.04 & 6.22 \\ 
  client & 0.09 & 0.09 & -0.000327 & 0.007314 & 0.000054 & 0.25 & 0.07 \\ 
  dealer & 0.06 & 0.13 & 0.062425 & 0.006891 & 0.003944 & 18.81 & 6.21 \\ 
  cook & 0.01 & 0.01 & 0.001548 & 0.002158 & 0.000007 & 0.48 & 8.03 \\ 
  thief & 0.02 & 0.04 & 0.017620 & 0.004041 & 0.000327 & 6.55 & 5.67 \\ 
  retired & 0.03 & 0.03 & 0.000793 & 0.004157 & 0.000018 & 0.27 & 1.49 \\ 
  homemakr & 0.06 & 0.06 & 0.003303 & 0.006172 & 0.000049 & 0.41 & 2.66 \\ 
  disabled & 0.04 & 0.06 & 0.015175 & 0.005137 & 0.000257 & 3.06 & 7.78 \\ 
  unemploy & 0.16 & 0.26 & 0.094812 & 0.009945 & 0.009088 & 25.32 & 9.26 \\ 
  homeless & 0.01 & 0.02 & 0.004798 & 0.002695 & 0.000030 & 1.16 & 5.03 \\ 
   \hline
\end{tabular}
\label{sbable}
\end{table}

%> print(xtable(data25a, caption='SB+ Design', label='sbplutable', digits=c(2,2,2,2,6,6,6,2,2)), include.rownames = F)
% latex table generated in R 3.4.4 by xtable 1.8-2 package
% Wed Oct  3 15:53:38 2018
\begin{table}[ht]
\centering
\caption{SB+ Design} 
\begin{tabular}{lrrrrrrr}
  \hline
SIMPLE & actual & E.est & bias & sd & mse & eff & rbias \\ 
  \hline
degree & 7.88 & 7.90 & 0.021062 & 0.302342 & 0.091854 & 1.00 & 1.00 \\ 
  nonwhite & 0.24 & 0.23 & -0.014272 & 0.022637 & 0.000716 & 1.00 & 1.00 \\ 
  female & 0.43 & 0.44 & 0.009500 & 0.028160 & 0.000883 & 1.00 & 1.00 \\ 
  worker & 0.05 & 0.06 & 0.004619 & 0.010680 & 0.000135 & 1.00 & 1.00 \\ 
  procurer & 0.02 & 0.02 & 0.000393 & 0.004503 & 0.000020 & 1.00 & 1.00 \\ 
  client & 0.09 & 0.08 & -0.012822 & 0.014368 & 0.000371 & 1.00 & 1.00 \\ 
  dealer & 0.06 & 0.06 & -0.002589 & 0.009764 & 0.000102 & 1.00 & 1.00 \\ 
  cook & 0.01 & 0.01 & -0.000378 & 0.004610 & 0.000021 & 1.00 & 1.00 \\ 
  thief & 0.02 & 0.02 & -0.001522 & 0.005890 & 0.000037 & 1.00 & 1.00 \\ 
  retired & 0.03 & 0.03 & -0.001919 & 0.008887 & 0.000083 & 1.00 & 1.00 \\ 
  homemakr & 0.06 & 0.06 & -0.000201 & 0.013147 & 0.000173 & 1.00 & 1.00 \\ 
  disabled & 0.04 & 0.04 & -0.000747 & 0.009763 & 0.000096 & 1.00 & 1.00 \\ 
  unemploy & 0.16 & 0.16 & 0.002081 & 0.018748 & 0.000356 & 1.00 & 1.00 \\ 
  homeless & 0.01 & 0.01 & -0.000520 & 0.005489 & 0.000030 & 1.00 & 1.00 \\ 
  \hline
VH & actual & E.est & bias & sd & mse & eff & rbias \\ 
  \hline
  degree & 7.88 & 5.19 & -2.696262 & 0.192625 & 7.306935 & 79.55 & 128.01 \\ 
  nonwhite & 0.24 & 0.27 & 0.033787 & 0.022481 & 0.001647 & 2.30 & 2.37 \\ 
  female & 0.43 & 0.39 & -0.040785 & 0.021736 & 0.002136 & 2.42 & 4.29 \\ 
  worker & 0.05 & 0.05 & -0.002137 & 0.009592 & 0.000097 & 0.71 & 0.46 \\ 
  procurer & 0.02 & 0.01 & -0.002142 & 0.003493 & 0.000017 & 0.82 & 5.46 \\ 
  client & 0.09 & 0.15 & 0.064674 & 0.019057 & 0.004546 & 12.26 & 5.04 \\ 
  dealer & 0.06 & 0.07 & 0.008032 & 0.010880 & 0.000183 & 1.79 & 3.10 \\ 
  cook & 0.01 & 0.01 & -0.001583 & 0.002833 & 0.000011 & 0.49 & 4.19 \\ 
  thief & 0.02 & 0.02 & 0.001872 & 0.006667 & 0.000048 & 1.30 & 1.23 \\ 
  retired & 0.03 & 0.03 & 0.004935 & 0.008288 & 0.000093 & 1.13 & 2.57 \\ 
  homemakr & 0.06 & 0.05 & -0.012813 & 0.008266 & 0.000232 & 1.34 & 63.75 \\ 
  disabled & 0.04 & 0.04 & -0.006037 & 0.007086 & 0.000087 & 0.90 & 8.08 \\ 
  unemploy & 0.16 & 0.13 & -0.029953 & 0.012773 & 0.001060 & 2.98 & 14.39 \\ 
  homeless & 0.01 & 0.01 & -0.000834 & 0.003959 & 0.000016 & 0.54 & 1.60 \\ 
  \hline
$\bar y$ & actual & E.est & bias & sd & mse & eff & rbias \\ 
  \hline
  degree & 7.88 & 14.23 & 6.350544 & 0.199966 & 40.369395 & 439.49 & 301.52 \\ 
  nonwhite & 0.24 & 0.30 & 0.058552 & 0.017339 & 0.003729 & 5.21 & 4.10 \\ 
  female & 0.43 & 0.46 & 0.024115 & 0.011285 & 0.000709 & 0.80 & 2.54 \\ 
  worker & 0.05 & 0.10 & 0.047787 & 0.005815 & 0.002317 & 17.12 & 10.35 \\ 
  procurer & 0.02 & 0.03 & 0.019401 & 0.003265 & 0.000387 & 18.95 & 49.43 \\ 
  client & 0.09 & 0.09 & -0.000469 & 0.007308 & 0.000054 & 0.14 & 0.04 \\ 
  dealer & 0.06 & 0.13 & 0.062280 & 0.006782 & 0.003925 & 38.46 & 24.05 \\ 
  cook & 0.01 & 0.01 & 0.001604 & 0.002308 & 0.000008 & 0.37 & 4.25 \\ 
  thief & 0.02 & 0.04 & 0.017384 & 0.004104 & 0.000319 & 8.62 & 11.42 \\ 
  retired & 0.03 & 0.03 & 0.000792 & 0.004025 & 0.000017 & 0.20 & 0.41 \\ 
  homemakr & 0.06 & 0.06 & 0.003582 & 0.005951 & 0.000048 & 0.28 & 17.82 \\ 
  disabled & 0.04 & 0.06 & 0.015048 & 0.005170 & 0.000253 & 2.64 & 20.14 \\ 
  unemploy & 0.16 & 0.26 & 0.095478 & 0.010154 & 0.009219 & 25.91 & 45.87 \\ 
  homeless & 0.01 & 0.02 & 0.004603 & 0.002667 & 0.000028 & 0.93 & 8.85 \\ 
   \hline
\end{tabular}
\label{sbplustable}
\end{table}

%rds
%print(xtable(data.frame(name=data25a$name[1:14], actual=data25a$actual[1:14], halfwidth=ci3rdata$half.width, coverage=ci3rdata$coverage), caption='RDS: Confidence Interval Coverage', label='RDScoverage'), include.rownames = FALSE)
% latex table generated in R 3.4.4 by xtable 1.8-2 package
% Sat Sep  1 17:33:05 2018
\begin{table}[ht]
\centering
\caption{RDS Design: Confidence Interval Coverage} 
\begin{tabular}{lrrr}
  \hline
name & actual & halfwidth & coverage \\ 
  \hline
degree & 7.88 & 0.58 & 0.80 \\ 
  nonwhite & 0.24 & 0.04 & 0.92 \\ 
  female & 0.43 & 0.05 & 0.97 \\ 
  worker & 0.05 & 0.02 & 0.95 \\ 
  procurer & 0.02 & 0.01 & 0.92 \\ 
  client & 0.09 & 0.03 & 0.94 \\ 
  dealer & 0.06 & 0.02 & 0.95 \\ 
  cook & 0.01 & 0.01 & 0.79 \\ 
  thief & 0.02 & 0.01 & 0.92 \\ 
  retired & 0.03 & 0.02 & 0.92 \\ 
  homemakr & 0.06 & 0.02 & 0.94 \\ 
  disabled & 0.04 & 0.02 & 0.94 \\ 
  unemploy & 0.16 & 0.03 & 0.95 \\ 
  homeless & 0.01 & 0.01 & 0.84 \\ 
   \hline
\end{tabular}
\label{RDScoverage}
\end{table}

% rdsplus
% print(xtable(data.frame(name=data25a$name[1:14], actual=data25a$actual[1:14], halfwidth=ci3adata$half.width, coverage=ci3adata$coverage), caption='RDS+: Confidence Interval Coverage', label='RDSPcoverage'), include.rownames = FALSE)
% latex table generated in R 3.4.4 by xtable 1.8-2 package
% Sun Sep  2 11:43:00 2018
\begin{table}[ht]
\centering
\caption{RDS+ Design: Confidence Interval Coverage} 
\begin{tabular}{lrrr}
  \hline
name & actual & halfwidth & coverage \\ 
  \hline
degree & 7.88 & 0.65 & 0.94 \\ 
  nonwhite & 0.24 & 0.04 & 0.85 \\ 
  female & 0.43 & 0.06 & 0.93 \\ 
  worker & 0.05 & 0.02 & 0.92 \\ 
  procurer & 0.02 & 0.01 & 0.76 \\ 
  client & 0.09 & 0.03 & 0.70 \\ 
  dealer & 0.06 & 0.02 & 0.79 \\ 
  cook & 0.01 & 0.01 & 0.64 \\ 
  thief & 0.02 & 0.01 & 0.67 \\ 
  retired & 0.03 & 0.02 & 0.86 \\ 
  homemakr & 0.06 & 0.03 & 0.92 \\ 
  disabled & 0.04 & 0.02 & 0.89 \\ 
  unemploy & 0.16 & 0.04 & 0.94 \\ 
  homeless & 0.01 & 0.01 & 0.68 \\ 
   \hline
\end{tabular}

\label{RDSPcoverage}
\end{table}

% sb
% print(xtable(data.frame(name=data25a$name[1:14], actual=data25a$actual[1:14], halfwidth=ci25rdata$half.width, coverage=ci25rdata$coverage), caption='SB: Confidence Interval Coverage', label='SBcoverage'), include.rownames = FALSE)
% latex table generated in R 3.4.4 by xtable 1.8-2 package
% Sun Sep  2 11:42:59 2018
\begin{table}[ht]
\centering
\caption{SB Design: Confidence Interval Coverage} 
\begin{tabular}{lrrr}
  \hline
name & actual & halfwidth & coverage \\ 
  \hline
degree & 7.88 & 0.59 & 0.72 \\ 
  nonwhite & 0.24 & 0.04 & 0.87 \\ 
  female & 0.43 & 0.05 & 0.97 \\ 
  worker & 0.05 & 0.02 & 0.95 \\ 
  procurer & 0.02 & 0.01 & 0.95 \\ 
  client & 0.09 & 0.03 & 0.96 \\ 
  dealer & 0.06 & 0.02 & 0.92 \\ 
  cook & 0.01 & 0.01 & 0.81 \\ 
  thief & 0.02 & 0.01 & 0.94 \\ 
  retired & 0.03 & 0.02 & 0.92 \\ 
  homemakr & 0.06 & 0.02 & 0.93 \\ 
  disabled & 0.04 & 0.02 & 0.94 \\ 
  unemploy & 0.16 & 0.03 & 0.94 \\ 
  homeless & 0.01 & 0.01 & 0.86 \\ 
   \hline
\end{tabular}

\label{SBcoverage}
\end{table}

% sbplus
% print(xtable(data.frame(name=data25a$name[1:14], actual=data25a$actual[1:14], halfwidth=ci25adata$half.width, coverage=ci25adata$coverage), caption='SB+: Confidence Interval Coverage', label='SBPcoverage'), include.rownames = FALSE)
% latex table generated in R 3.4.4 by xtable 1.8-2 package
% Sat Sep  1 17:54:29 2018
\begin{table}[ht]
\centering
\caption{SB+ Design: Confidence Interval Coverage} 
\begin{tabular}{lrrr}
  \hline
name & actual & halfwidth & coverage \\ 
  \hline
degree & 7.88 & 0.66 & 0.96 \\ 
  nonwhite & 0.24 & 0.04 & 0.89 \\ 
  female & 0.43 & 0.06 & 0.94 \\ 
  worker & 0.05 & 0.02 & 0.95 \\ 
  procurer & 0.02 & 0.01 & 0.87 \\ 
  client & 0.09 & 0.03 & 0.76 \\ 
  dealer & 0.06 & 0.02 & 0.85 \\ 
  cook & 0.01 & 0.01 & 0.65 \\ 
  thief & 0.02 & 0.01 & 0.73 \\ 
  retired & 0.03 & 0.02 & 0.83 \\ 
  homemakr & 0.06 & 0.03 & 0.92 \\ 
  disabled & 0.04 & 0.02 & 0.90 \\ 
  unemploy & 0.16 & 0.04 & 0.95 \\ 
  homeless & 0.01 & 0.01 & 0.67 \\ 
   \hline
\end{tabular}
\label{SBPcoverage}
\end{table}

\pagebreak

% mean degree, all designs

\begin{figure}
\begin{center}
  \includegraphics[width = 0.7\textwidth, angle =
  0]{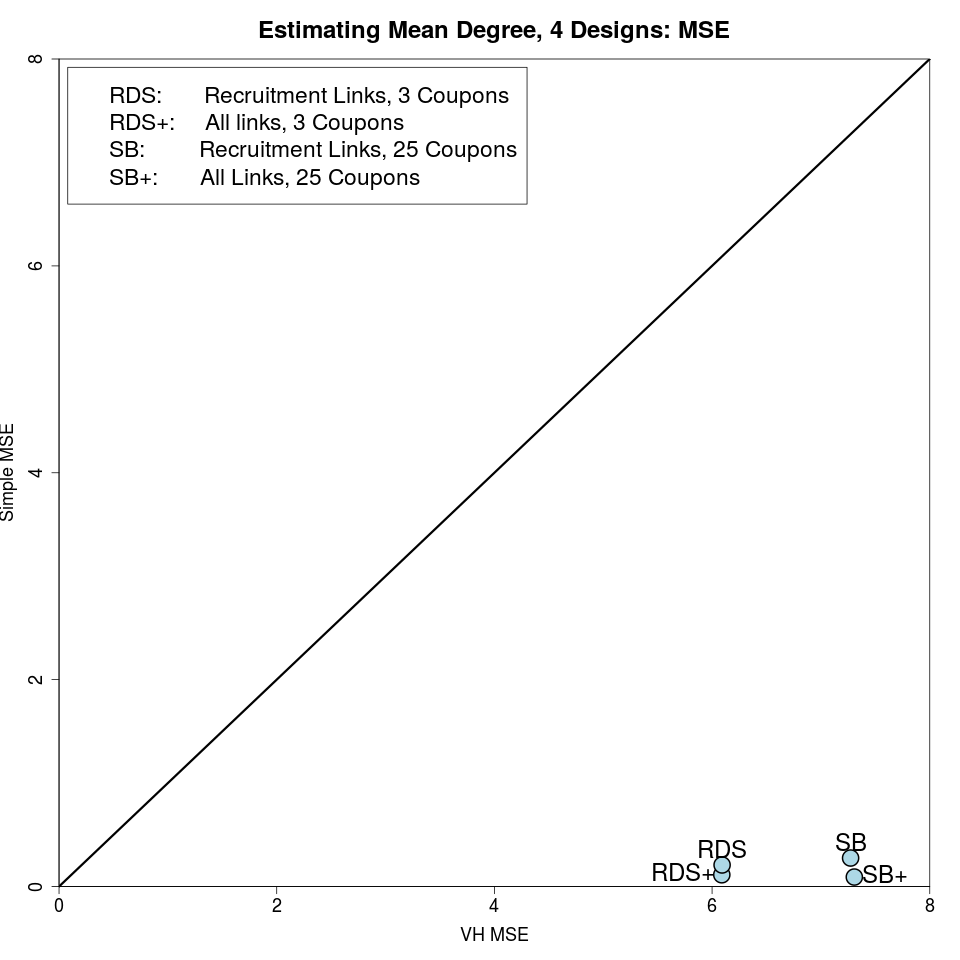}
\caption{MSE for mean degree, all designs}
\label{figmsedegree}
\end{center} 
\end{figure}

\begin{figure}
\begin{center}
  \includegraphics[width = 0.7\textwidth, angle =
  0]{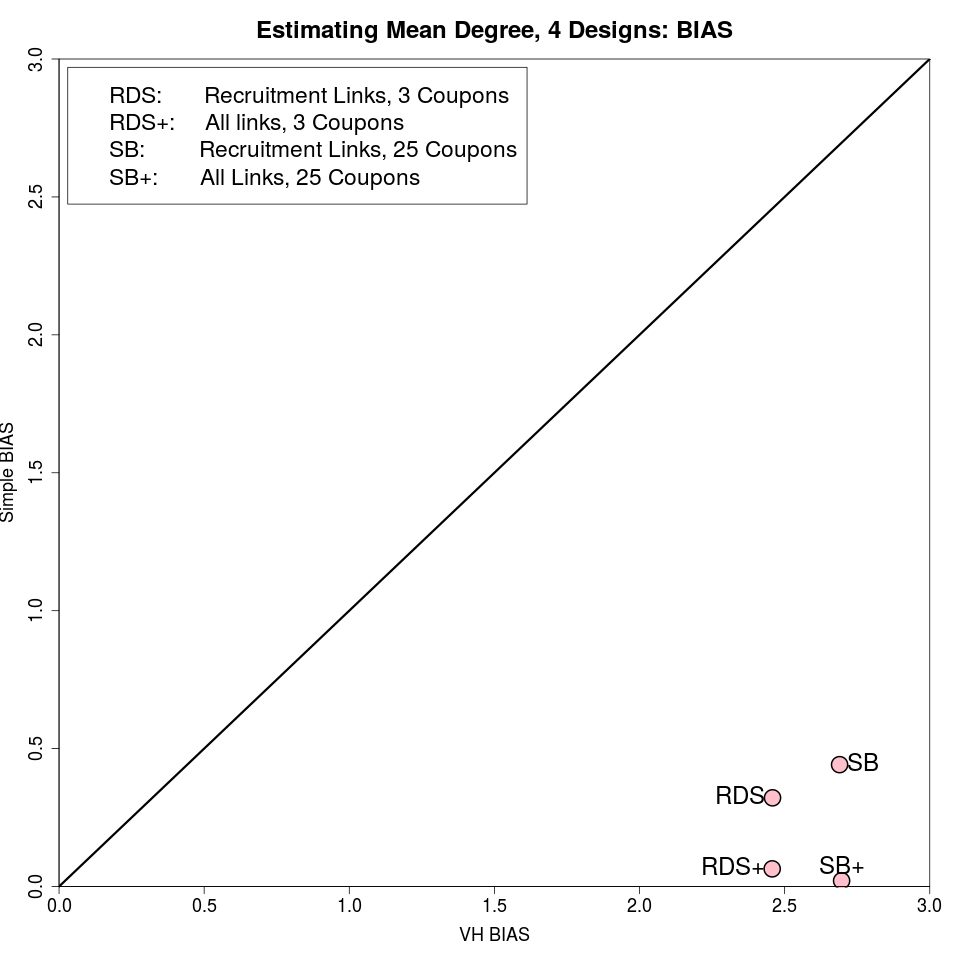}
\caption{Bias for mean degree, all designs}
\label{figbiasdegree}
\end{center} 
\end{figure}

%rds mse

\begin{figure}
\begin{center}
  \includegraphics[width = 0.7\textwidth, angle =
  0]{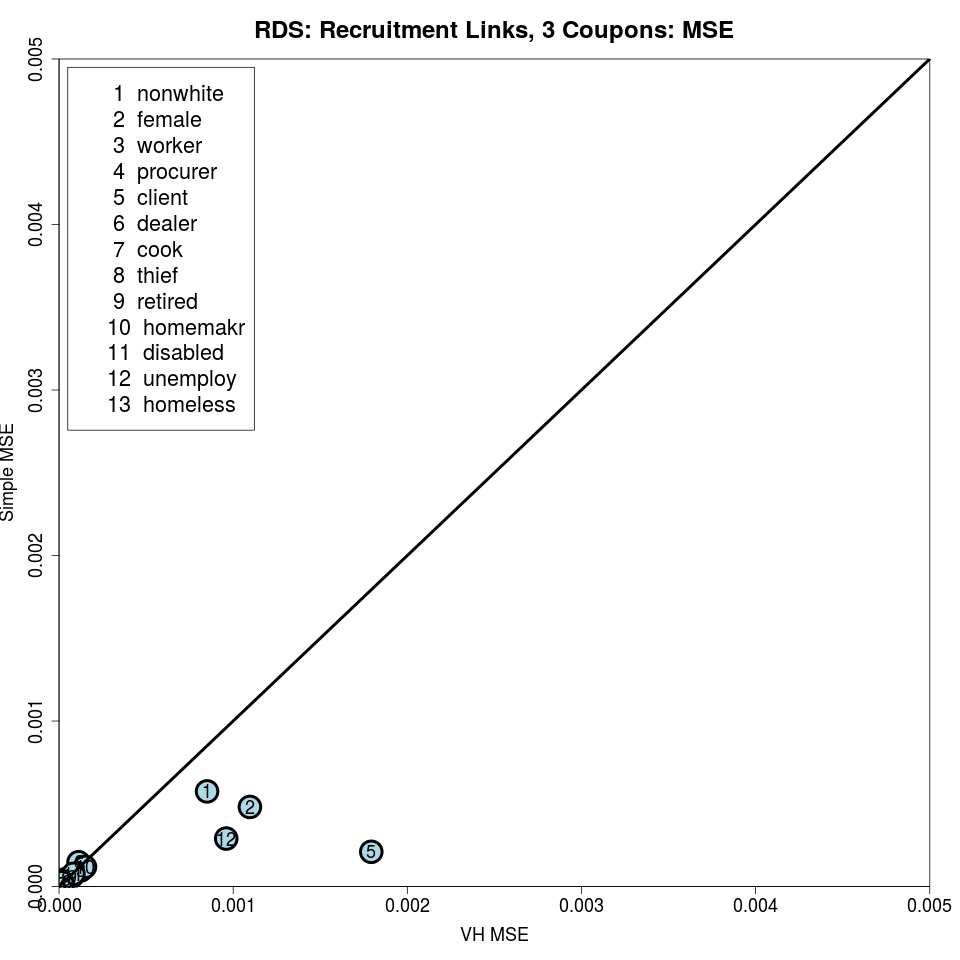}
\caption{MSE, RDS design}
\label{figmserds}
\end{center} 
\end{figure}

% rdsplus mse 

\begin{figure}
\begin{center}
  \includegraphics[width = 0.7\textwidth, angle =
  0]{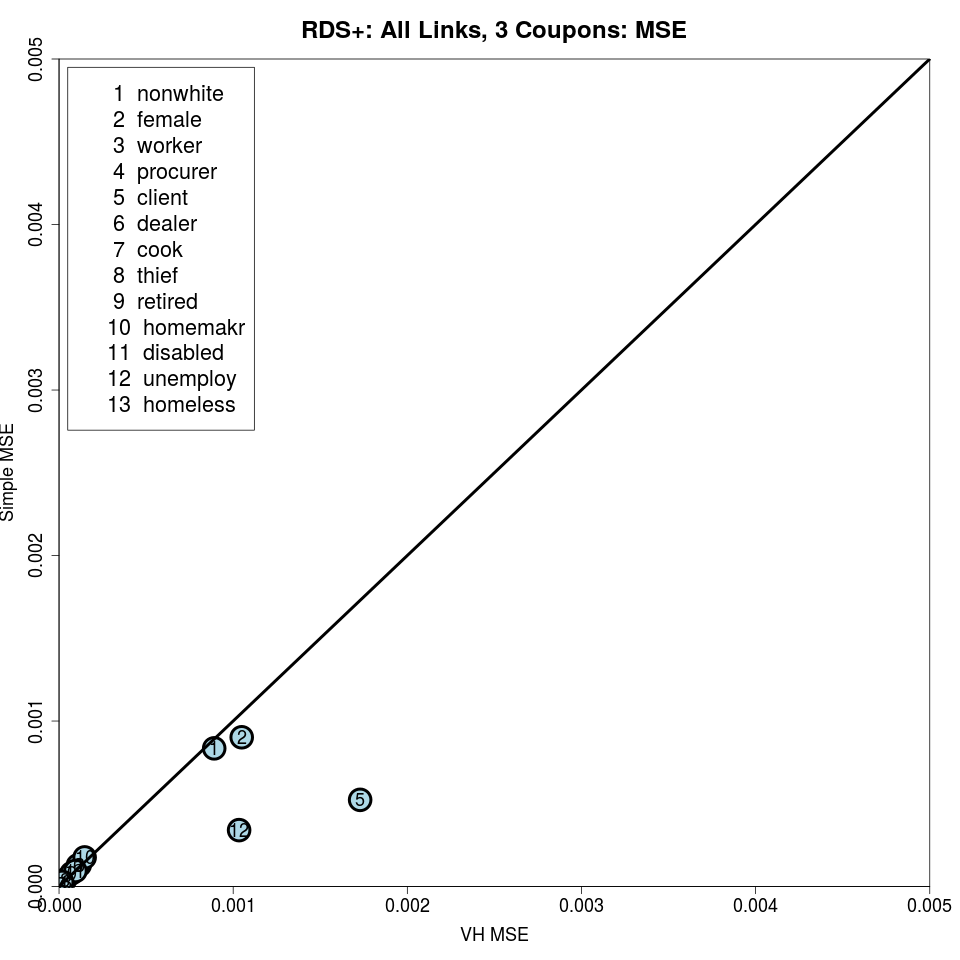}
\caption{MSE, RDS+ design}
\label{figmserdsplus}
\end{center} 
\end{figure}

%sb mse

\begin{figure}
\begin{center}
  \includegraphics[width = 0.7\textwidth, angle =
  0]{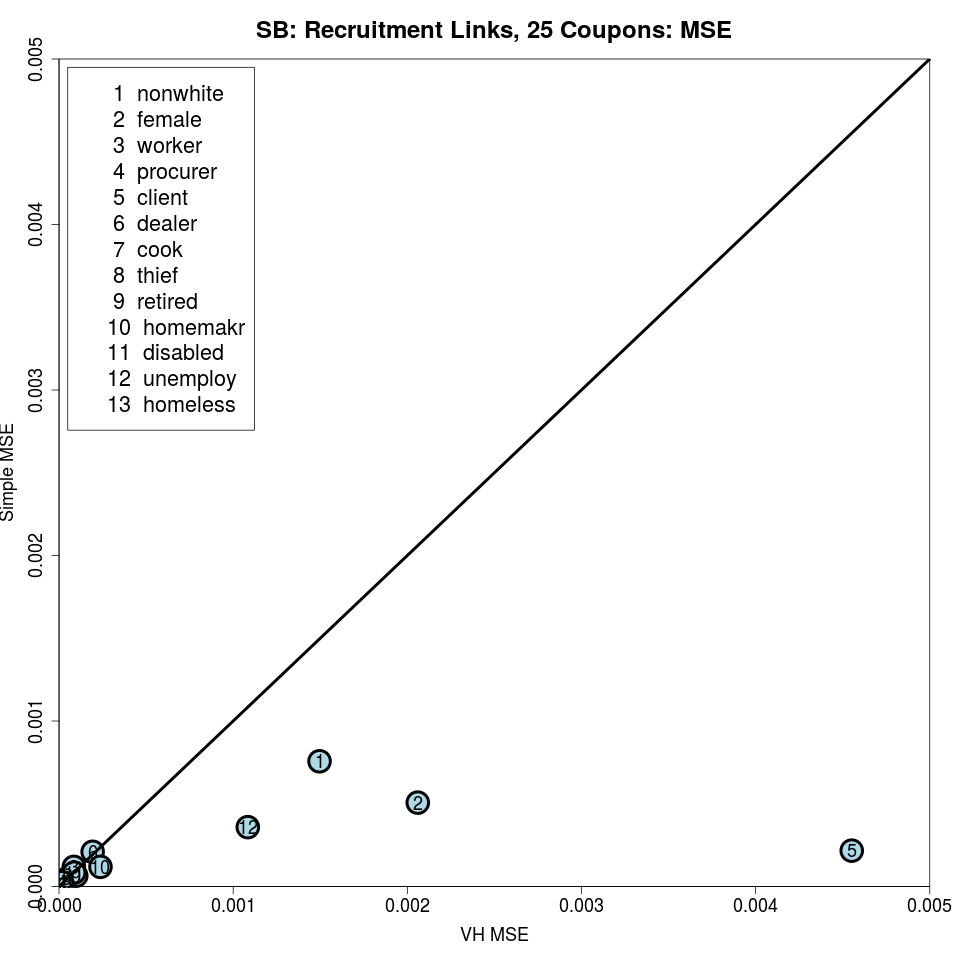}
\caption{MSE, SB design}
\label{figmsesb}
\end{center} 
\end{figure}

% sbplus mse 

\begin{figure}
\begin{center}
  \includegraphics[width = 0.7\textwidth, angle =
  0]{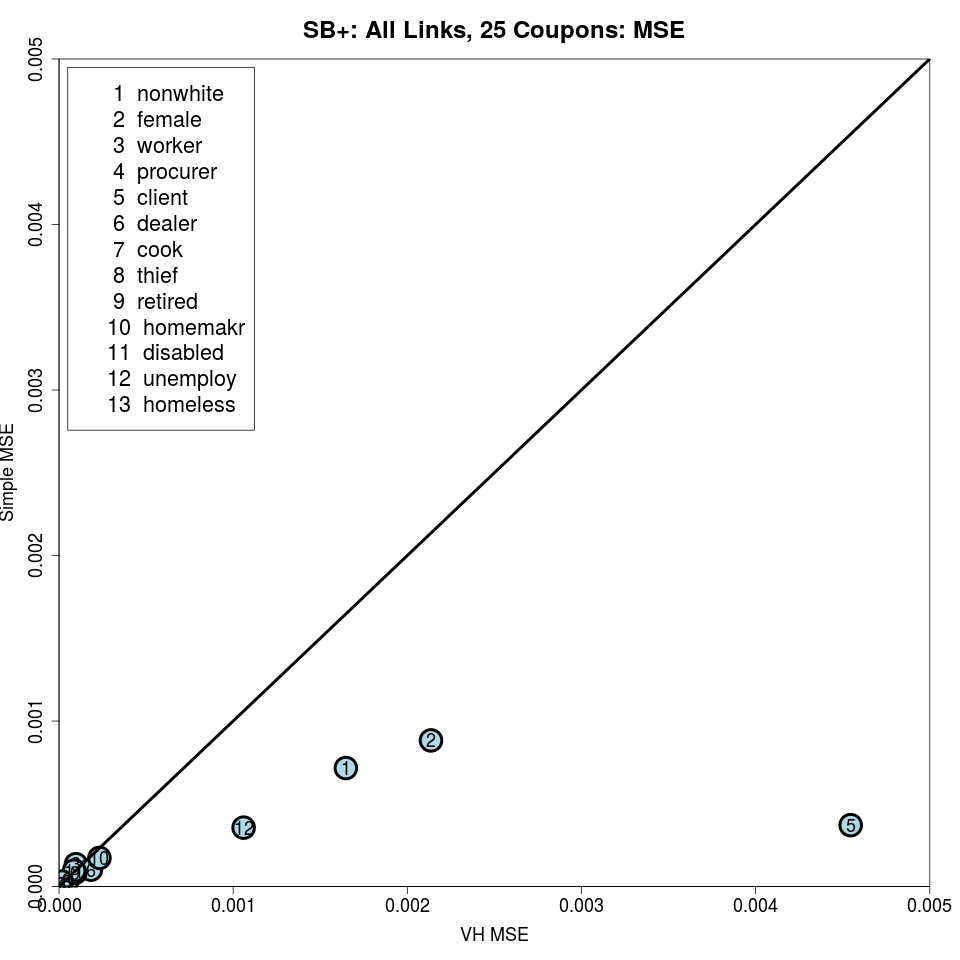}
\caption{MSE, SB+ design}
\label{figmsesbplus}
\end{center} 
\end{figure}

% rds bias

\begin{figure}
\begin{center}
  \includegraphics[width = 0.7\textwidth, angle =
  0]{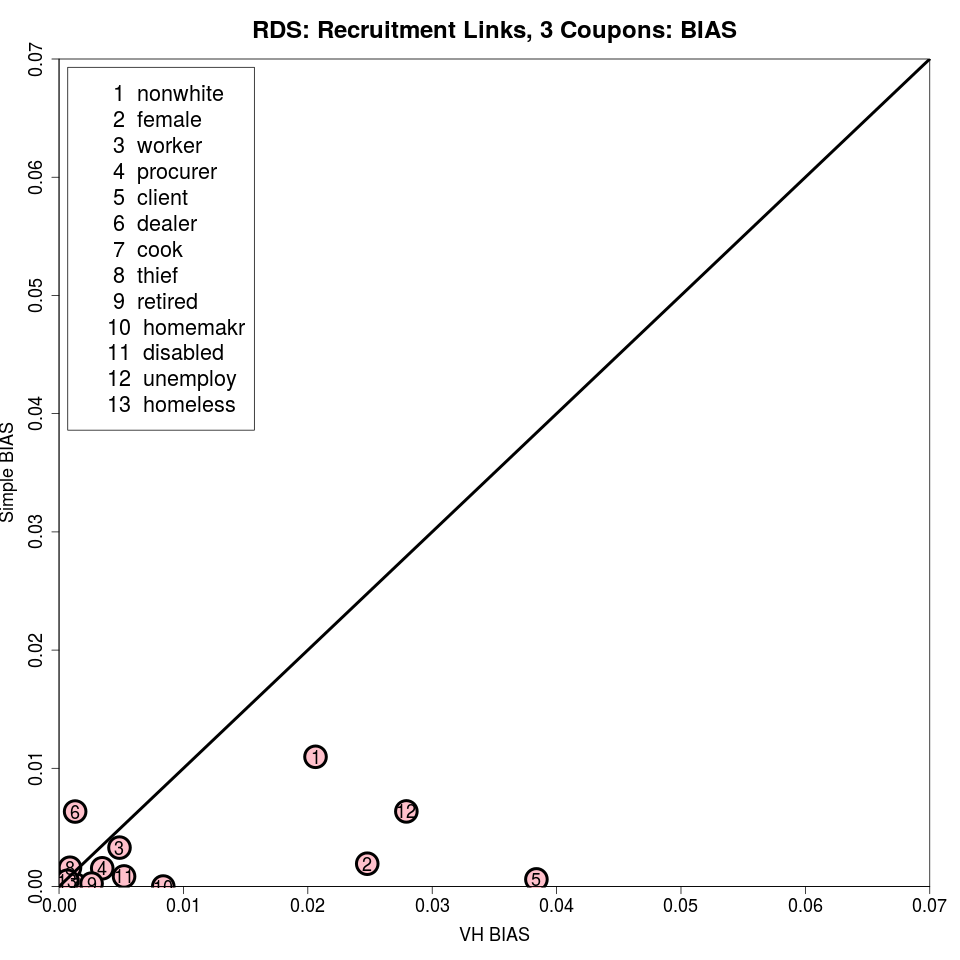}
\caption{Bias, RDS design}
\label{figbiasrds}
\end{center} 
\end{figure}

%rdsplus bias

\begin{figure}
\begin{center}
  \includegraphics[width = 0.7\textwidth, angle =
  0]{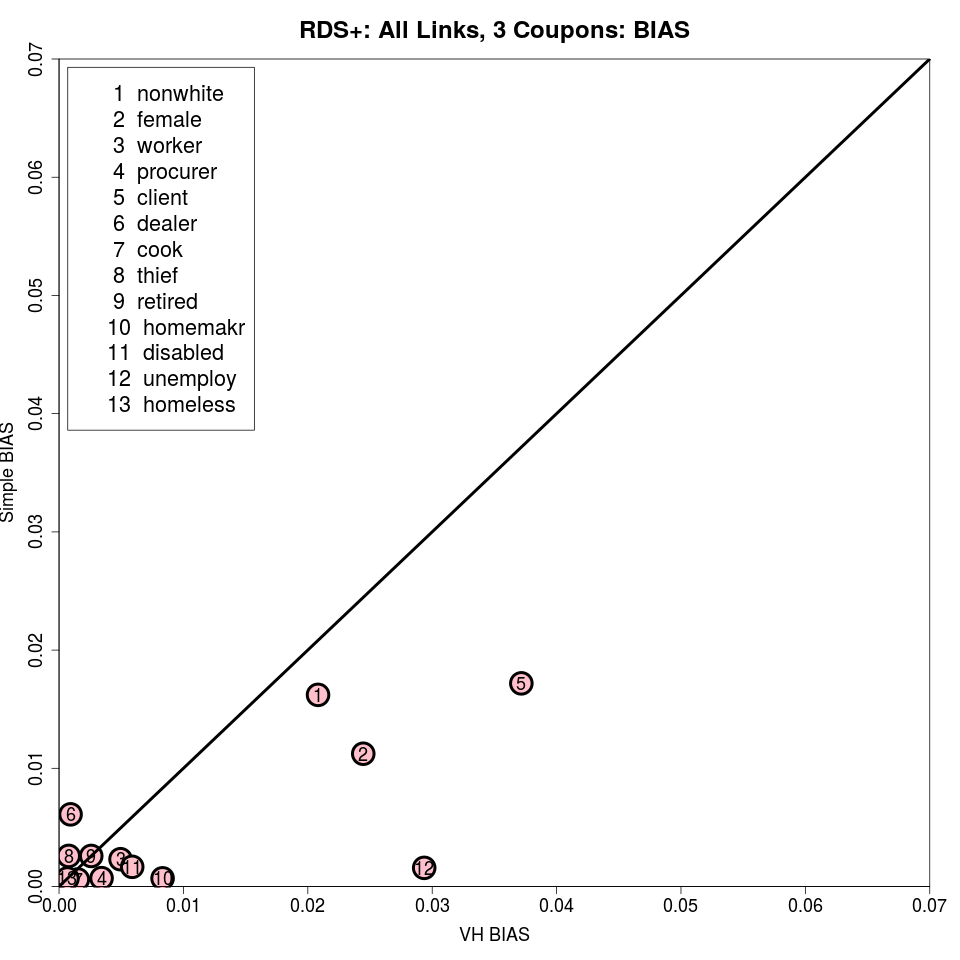}
\caption{Bias, RDS+ design}
\label{figbiasrdsplus}
\end{center} 
\end{figure}

% sb bias

\begin{figure}
\begin{center}
  \includegraphics[width = 0.7\textwidth, angle =
  0]{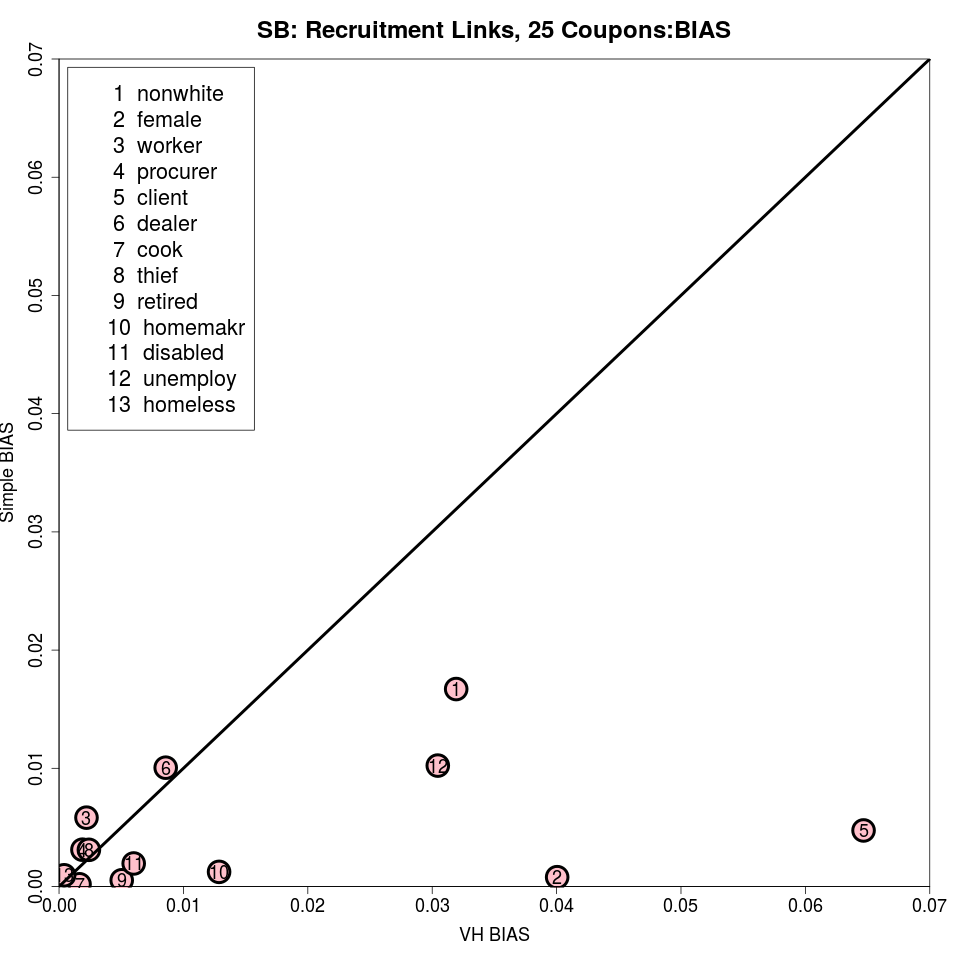}
\caption{Bias, SB design}
\label{figbiassb}
\end{center} 
\end{figure}

% sbplus bias

\begin{figure}
\begin{center}
  \includegraphics[width = 0.7\textwidth, angle =
  0]{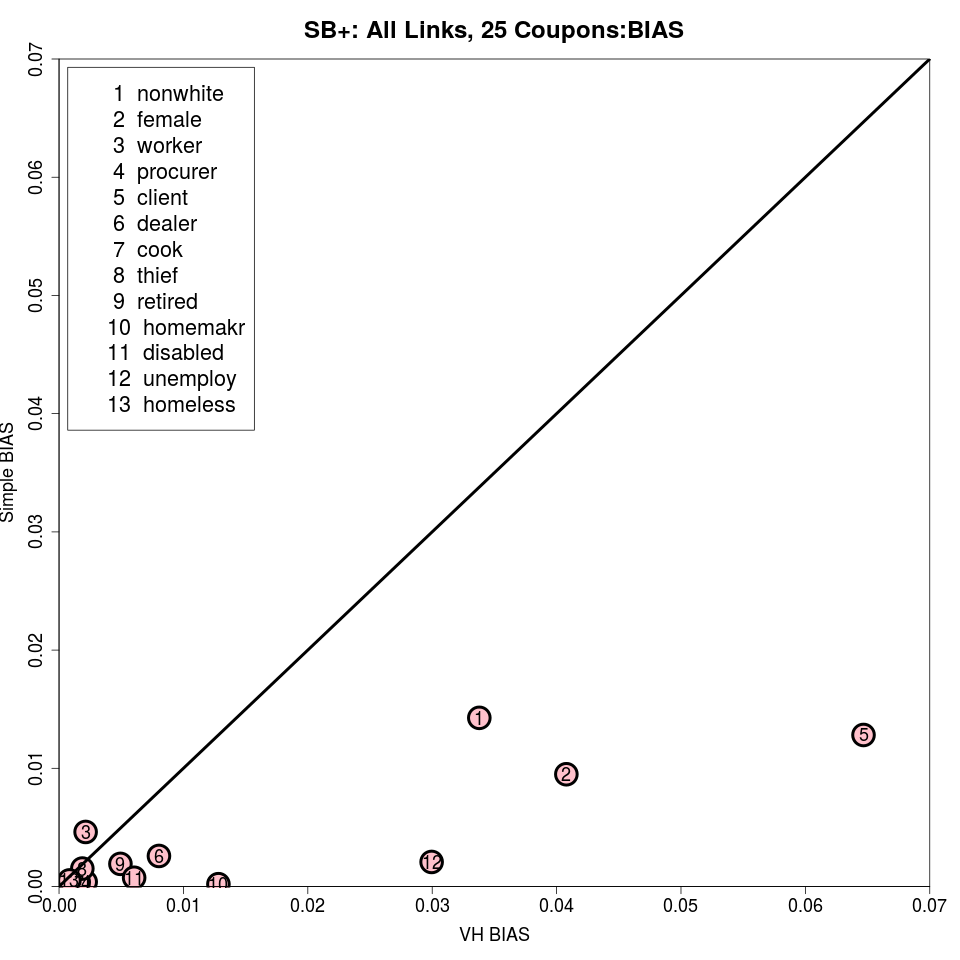}
\caption{Bias, RDS+ design}
\label{figbiassbplus}
\end{center} 
\end{figure}

\end{document}